\documentclass[a4paper,12pt]{article}
\pdfoutput=1
\usepackage{graphicx, rotating,amssymb,tabularx,colortbl}

\ifx\pdfoutput\undefined
\usepackage[dvips,bookmarks]{hyperref}	
\else
\usepackage{hyperref}	
\fi
\hypersetup{colorlinks,bookmarksopen,bookmarksnumbered,citecolor=verdes,
linkcolor=blus,pdfstartview=FitH,urlcolor=rossos}

\def\hhref#1{\href{http://arxiv.org/abs/#1}{#1}} 

\usepackage{multicol}
\usepackage{color}

\definecolor{rosso}{cmyk}{0,1,1,0.4}
\definecolor{rossos}{cmyk}{0,1,1,0.55}
\definecolor{rossoc}{cmyk}{0,1,1,0.2}
\definecolor{blu}{cmyk}{1,1,0,0.3}
\definecolor{blus}{cmyk}{1,1,0,0.6}
\definecolor{bluc}{cmyk}{1,1,0,0.1}
\definecolor{verde}{cmyk}{0.92,0,0.59,0.25}
\definecolor{verdec}{cmyk}{0.92,0,0.59,0.15}
\definecolor{verdes}{cmyk}{0.92,0,0.59,0.4}
\definecolor{verdino}{cmyk}{0.50,0,0.99,0.00}
\definecolor{grigio}{cmyk}{0,0,0,0.4}
\definecolor{grigioc}{cmyk}{0,0,0,0.3}
\definecolor{giallo}{cmyk}{0,0,0.8,0}

\font\tenrsfs=rsfs10 at 12pt
\font\sevenrsfs=rsfs7
\font\fiversfs=rsfs5
\newfam\rsfsfam
\textfont\rsfsfam=\tenrsfs
\scriptfont\rsfsfam=\sevenrsfs
\scriptscriptfont\rsfsfam=\fiversfs
\def\mathscr#1{{\fam\rsfsfam\relax#1}}
\def\Lag{\mathscr{L}}

\newcommand{\fig}[1]{~\ref{fig:#1}}
\newcommand{\eq}[1]{~{\rm (\ref{eq:#1})}}

\newcommand{\GeV}{\,{\rm GeV}}
\newcommand{\TeV}{\,{\rm TeV}}
\newcommand{\eV}{\,{\rm eV}}

\def\circa#1{\,\raise.3ex\hbox{$#1$\kern-.75em\lower1ex\hbox{$\sim$}}\,}
\newcommand{\nubarnu}{\raisebox{1ex}{\hbox{\tiny(}}\overline\nu\raisebox{1ex}{\hbox{\tiny)}}\hspace{-0.5ex}}

\newcommand{\DM}{{\rm DM}}

\newcommand{\PR}{Phys. Rev.}

\newcommand{\beq}{\begin{equation}}
\newcommand{\eeq}{\end{equation}}
\newcommand{\MeV}{\,{\rm MeV}}
\def\circa#1{\,\raise.3ex\hbox{$#1$\kern-.75em\lower1ex\hbox{$\sim$}}\,}
\makeatletter

\def\art{\@ifnextchar[{\eart}{\oart}}
\def\eart[#1]#2#3#4#5#6{{\rm #2}, {#3 #4} {\rm (#6) #5} [{\hhref{#1}}]}
\def\hepart[#1]#2{{\rm #2, \hhref{#1}}}
\newcommand{\oart}[5]{{\rm #1}, {#2 #3} {\rm (#5) #4}}

\newcounter{alphaequation}[equation]
\def\thealphaequation{\theequation\hbox to
0.6em{\hfil\alph{alphaequation}\hfil}}
\def\eqnsystem#1{
\def\@eqnnum{{\rm (\thealphaequation)}}
\def\@@eqncr{\let\@tempa\relax \ifcase\@eqcnt \def\@tempa{& & &} \or
  \def\@tempa{& &}\or \def\@tempa{&}\fi\@tempa
  \if@eqnsw\@eqnnum\refstepcounter{alphaequation}\fi
\global\@eqnswtrue\global\@eqcnt=0\cr}
\refstepcounter{equation} \let\@currentlabel\theequation \def\@tempb{#1}
\ifx\@tempb\empty\else\label{#1}\fi
\refstepcounter{alphaequation}
\let\@currentlabel\thealphaequation
\global\@eqnswtrue\global\@eqcnt=0 \tabskip\@centering\let\\=\@eqncr
$$\halign to \displaywidth\bgroup \@eqnsel\hskip\@centering
$\displaystyle\tabskip\z@{##}$&\global\@eqcnt\@ne
\hskip2\arraycolsep\hfil${##}$\hfil& \global\@eqcnt\tw@\hskip2\arraycolsep
$\displaystyle\tabskip\z@{##}$\hfil
\tabskip\@centering&\llap{##}\tabskip\z@\cr}
\def\endeqnsystem{\@@eqncr\egroup$$\global\@ignoretrue} \makeatother

\newcommand{\SU}{\rm SU}

\begin{document}
\begin{center}
{IFUP-TH/2009-04}
{ \hfill SACLAY--T09/010}
\color{black}
\vspace{1cm}

{\Huge\bf Minimal Dark Matter:\\[3mm] model and results}

\medskip
\bigskip\color{black}\vspace{0.6cm}

{
{\large\bf Marco Cirelli}$^a$,
{\large\bf Alessandro Strumia}$^b$
}
\\[7mm]
{\it $^a$ Institut de Physique Th\'eorique, CNRS, URA 2306 \& CEA/Saclay,\\ 
	F-91191 Gif-sur-Yvette, France}\\[3mm]
{\it $^b$ Dipartimento di Fisica dell'Universit\`a di Pisa and INFN, Italia}\\[3mm]
\end{center}

\bigskip

\centerline{\large\bf Abstract}
\begin{quote}
\color{black}\large
We recap the main features of Minimal Dark Matter (MDM) and assess its status in the light of the recent experimental data. The theory selects an electroweak 5-plet with hypercharge $Y=0$ as a fully successful DM candidate, automatically stable against decay and with no free parameters: DM is a fermion with a $9.6\TeV$ mass.
The direct detection cross-section, predicted to be $10^{-44}$ cm$^2$, is within reach of next-generation experiments. DM is accompanied by a charged fermion 166 MeV heavier: we discuss how it might manifest.
Thanks to an electroweak Sommerfeld enhancement of more than 2 orders of magnitude, DM annihilations into $W^+W^-$ give, in presence of a modest  astrophysical boost factor, an $e^+$ flux compatible with the PAMELA excess (but not with the ATIC hint for a peak: MDM instead predicts a quasi-power-law spectrum), a $\bar p$ flux concentrated at energies above 100 GeV, and to photon fluxes comparable with present limits, depending on the DM density profile.
\end{quote}


\section{Introduction}

The quest for the identification of the missing mass of the universe has been with us since many decades now~\cite{reviews}. While explanations in terms of modifications of Newtonian gravity or General Relativity become more and more contrived, evidence for the particle nature of such Dark Matter (DM) now comes from many astrophysical and cosmological observations. 
Non-baryonic new particles that may fulfill the r\^ole of DM have emerged in the latest decades within many 
Beyond the Standard Model (SM) theories, most notably supersymmetry. 
These constructions try to naturally explain the hierarchy between the ElectroWeak (EW) scale and the Planck scale and, in doing so, introduce a host of new particles with EW masses and interactions. 
Some of these particles can be good DM candidates (e.g.\ the lightest neutralino).

\smallskip

However these approaches to the solution of the DM problem, while still the most popular, start facing a sort of `impasse': (i) The expected new physics at the EW scale has not appeared so far at collider experiments:
the simplest solutions to the hierarchy problem start needing uncomfortably high fine-tunings of the their unknown parameters~\cite{FT}.
(ii) The presence of a number of unknown parameters (e.g.\ all sparticle masses) obscures the phenomenology of the DM candidates.
(iii) The stability of the DM candidates is usually the result of extra features introduced by hand (e.g.\  $R$-parity in supersymmetry), most often also necessary to recover many good properties of the SM that are lost in these extensions (automatic conservation of baryon number, lepton number, etc).

The Minimal Dark Matter (MDM) proposal~\cite{MDM} pursues therefore a different and somewhat opposite direction: focussing on the Dark Matter problem only, we add to the SM the minimal amount of new physics (just one extra EW multiplet ${\cal X}$) and search for the minimal assignments of its quantum numbers (spin, isospin and hypercharge) that make it a good Dark Matter candidate without ruining the positive features of the SM. No ad hoc extra features are introduced: the stability of the successful candidates is guaranteed by the SM gauge symmetry and by renormalizability. Moreover, due to its minimality, the theory is remarkably predictive: no free parameters are present and therefore the phenomenological signatures can be univocally calculated, e.g. at colliders and for direct/indirect detection, up to the astrophysical and cosmological uncertainties. 

\bigskip

The rest of the paper is organized as follows. In Sec.\ref{model} we review the theoretical aspects of MDM model. We follow a constructive approach by scanning all the possible choices of quantum numbers and selecting the successful DM candidates: insisting on full minimality we will see that consistency and phenomenological constraints reject most of the candidates and actually individuate only one (the fermionic SU(2) quintuplet with hypercharge $Y=0$), of which we will study the phenomenology in the following sections. Note that since other candidates can be re-allowed by relaxing the request for full minimality (as we discuss in Appendix B), the formulae and most of the phenomenological analysis are given in their most general form and can easily be adapted to these other candidates.
In Sec.\ref{relic} we present the computation of the cosmological relic density of MDM~\cite{MDM2}: imposing the match with the measured density determines the DM mass. We also discuss the small mass splitting introduced by loop corrections.
In Sec.\ref{direct} we review the direct detection signatures of the candidate. In Sec.\ref{indirect} we focus on the indirect detection signatures, comparing the predictions (\cite{MDM3}) with the recent data from satellite and balloon experiments: this promises to be the most interesting and stringent avenue of confrontation with data~\cite{MDMidm08}. In Sec.\ref{collider} we review what could be the signatures at colliders. 
Finally, in Sec.\ref{other} we discuss some other possible phenomenological features of the model. 
In Appendix A we briefly review the Sommerfeld enhancement of the annihilation cross section, discussed in more detail in~\cite{MDM2, MDM3}. In Appendix B we list the options that open up for model building if the requirement of full minimality is relaxed.
An executive summary of positive and negative features of the model, as well as an outlook towards the implications of future results, is given in Sec.\ref{conclusions}.

\section{Construction of the Minimal Dark Matter model}
\label{model}

\begin{table}[t]
\center
\begin{tabularx}{0.721 \textwidth}{|ccc|c||c|c|}
\hline
\multicolumn{3}{|c|}{\hbox{Quantum numbers}}
&\hbox{DM can} & \hbox{DD}& \hbox{Stable?}  \\
 $\SU(2)_L$ & ${\rm U}(1)_Y$ & \hbox{Spin} & \hbox{decay into} & \hbox{bound?} & \\
 \hline
 \hline
 2 & 1/2 & $S$ & $EL$ &  $\times$ & $\times$ \\ 
2 & 1/2 & $F$ & $EH$ & $\times$ & $\times$ \\ 
\hline
3 & 0 & $S$ & $HH^*$ & $\surd$ & $\times$ \\ 
3 & 0 & $F$ & $LH$ & $\surd$ & $\times$ \\ 
3 & 1 & $S$ & $HH,LL$ & $\times$ & $\times$\\ 
3 & 1 & $F$ & $LH$ & $\times$ &$\times$ \\ 
\hline
4 & 1/2 & $S$ & $HHH^*$ & $\times$ & $\times$\\ 
4 & 1/2 & $F$ & $(LHH^*)$ & $\times$ &$\times$ \\ 
4 & 3/2 & $S$ & $HHH$ & $\times$ & $\times$ \\ 
4 & 3/2 & $F$ & $(LHH)$ & $\times$ & $\times$ \\ 
\hline
5 & 0 & $S$ & $(HHH^*H^*)$ & $\surd$ & $\times$ \\ 
\rowcolor{verdino} 5 & 0 & $F$ & $-$ & $\surd$ & $\surd$ \\ 
5 & 1 & $S$ & $(HH^*H^*H^*)$ & $\times$ & $\times$ \\ 
5 & 1 & $F$ & $-$ & $\times$ & $\surd$ \\ 
5 & 2 & $S$ & $(H^*H^*H^*H^*)$ & $\times$ & $\times$ \\ 
5 & 2 & $F$ & $-$ & $\times$ & $\surd$ \\ 
\hline
6 & 1/2, 3/2, 5/2  & $S$ & $-$ & $\times$ & $\surd$ \\  
\hline
\rowcolor{giallo} 7 & 0 & $S$ & $-$ & $\surd$ & $\surd$ \\  
\hline
8 & 1/2, 3/2 $\ldots$  & $S$ & $-$ & $\times$ & $\surd$ \\  
\hline
\end{tabularx}
 \caption{\em\label{tab:1} {\bf Book-keeping of the possible Minimal DM candidates and selection of successful ones}.
Quantum numbers are listed in the first 3 columns. The 4th column indicates some decay modes into SM particles; modes listed in parenthesis correspond to dimension 5 operators. Candidates with $Y\neq 0$ are excluded by Direct Detection (DD) searches (unless appropriate non-minimalities are introduced, see App.B), as indicated in the 5th column. Candidates with an open decay channel are excluded (unless some other non-minimalities are introduced, see App.B), as indicated in the 6th column. Note that, for simplicity, the possibilities concerning the $SU_L(2)$ 6- and 8-plet are only sketched. Analogously, for the 7-plet we list the only interesting candidate. At the end of the game, the fully successful candidates are indicated by the shaded background: the fermionic 5-plet with $Y=0$ and the scalar 7-plet with $Y=0$. As for the latter non-minimal scalar quartic couplings are generically present (see Appendix B), the former is overall preferred and we will refer to it as {\em the} MDM candidate in the paper.}
\end{table}

The MDM model is constructed  by simply adding on top of the Standard Model a single fermionic or scalar multiplet ${\cal X}$ charged under the usual SM $SU_L(2)\times U_Y(1)$ electroweak interactions (that is: a WIMP; it is assumed not to be charged under $SU_c(3)$ strong interactions as the bounds are strong on this possibility~\cite{strong}). 
Its conjugate  $\bar{\cal X}$ belongs to the same representation, so that the theory is vector-like with respect to $SU_L(2)$ and anomaly-free.
The Lagrangian is `minimal':
\beq
\label{eq:lagrangian}
\mathscr{L} = \mathscr{L}_{\rm SM} + \frac{1}{2}
\left\{\begin{array}{ll}
 \bar{\cal X} (i D\hspace{-1.4ex}/\hspace{0.5ex}+M) {\cal X} & \hbox{for fermionic ${\cal X}$}\\
|D_\mu {\cal X}|^2 - M^2 |{\cal X}|^2& \hbox{for scalar ${\cal X}$}
\end{array}\right.
\eeq
The gauge-covariant derivative $D_\mu $ contains the known electroweak gauge couplings to the vectors bosons of the SM ($Z$, $W^\pm$ and $\gamma$) and $M$ is a tree level mass term (the only free parameter of the theory).
A host of additional term (such as Yukawa couplings with SM fields) would in principle be present, but for successful candidates they will be forbidden by gauge and Lorentz invariance, as detailed below. 

${\cal X}$ is fully determined by the assignments of its quantum numbers under the gauge group: the number of its $SU(2)_L$ components, $n=\{2,3,4,5,\ldots\}$ and the hypercharge $Y$. In Table~\ref{tab:1} we list all the potentially successful combinations, as we now proceed to discuss.

\smallskip

For a given assignment of $n$ (first column of the table) there are a few choices of the hypercharge $Y$ such that one component of the ${\cal X}$ multiplet has electric charge $Q=T_3+Y=0$ (where $T_3$ is the usual `diagonal' generator of $\SU(2)_L$), as needed for a DM candidate.
For instance, for the doublet $n=2$, since $T_3 = \pm 1/2$, the only possibility is $Y=\mp 1/2$.
For $n=5$ one can have $Y = \{ 0, \pm 1, \pm 2\}$, and so on.
We do not consider the case of the $n=1$ singlet: lacking gauge interactions, even if it is ever produced in the Early Universe it could not annihilate and remain with the correct relic amount by means of the standard freeze-out mechanism.\footnote{We discuss the case of a scalar singlet with non-minimal additional interactions in Appendix B.}\\
The list of possible candidates has to stop at $n \le 5~ (8)$ for fermions (scalars) because larger multiplets would accelerate the running of the $SU(2)_L$ coupling $g_2$: demanding that the perturbativity of $\alpha_2^{-1}(E') = \alpha_2^{-1}(M) - (b_2/2\pi)\ln E'/M$ is mantained all the way up to  $E'\sim M_{\rm Pl}$ (since the Planck scale $M_{\rm Pl}$ is the cutoff scale of the theory) imposes the bound. In this formula $b_2 = -19/6 + c\ g_{\cal X} (n^2-1)/36$ with $c=1$ for fermions, $c=1/4$ for scalars and $g_{\cal X}$ is the number of degrees of freedom in the multiplet.

\smallskip

In this list of candidates, those with $Y\neq 0$ have vector-like interactions with the $Z$ boson that produce a tree-level spin-independent elastic cross sections 
\beq 
\label{eq:direct}
\sigma({\rm DM} \, {\cal N}\to {\rm DM}\,{\cal N}) =c\frac{G_{\rm F}^2M_{\cal N}^2}{2\pi} Y^2(N - (1-4s_{\rm W}^2) Z)^2 \eeq
where $c=1$ for fermionic DM and $c=4$ for scalar DM~\cite{GoodWit};
$Z$ and $N$ are the number of protons and of neutrons in the target nucleus with mass $M_{\cal N}$ (we are assuming $M\gg M_{\cal N}$). This elastic cross section is $2\div 3$ orders of magnitude above the present bounds~\cite{CDMS, Xenon} from direct detection searches. 
Unless minimality is abandoned in an appropriate way (that we discuss in Appendix B), such MDM candidates are therefore excluded and we will focus in the following on those with $Y=0$.

\smallskip

Next we need to inspect which of the remaining candidates are stable against decay into SM particles. The fourth column of Table~\ref{tab:1} shows some possible decay operators for each case. For instance, the fermionic 3-plet with hypercharge $Y=0$ would couple through a Yukawa operator ${\cal X} L H$ with a SM lepton doublet $L$ and a Higgs field $H$ and decay in a very short time. This is not a viable DM candidate, unless the operator is eliminated by some ad hoc symmetry (see again Appendix B).
For another instance, the scalar 5-plet with $Y=0$ would couple to four Higgs fields with a dimension 5 operator ${\cal X} HHH^*H^*/M_{\rm Pl}$, suppressed by one power of the Planck scale. Despite the suppression, the resulting typical life-time $\tau \sim M_{\rm Pl}^2\, {\rm TeV}^{-3}$ is shorter than the age of the universe, so that this is not a viable DM candidate.\\
Now, the crucial observation is that, given the known SM particle content, the large $n$ multiplets cannot couple to SM fields and are therefore automatically stable DM candidates. This is the same reason why known massive stable particles (like the proton) are stable: decay modes consistent with renormalizability and gauge symmetry do not exist. 
In other words, for these candidates DM stability is explained by an `accidental symmetry', like proton stability.
Among the candidates that survived all the previous constraints, only two possibilities then emerge:
a $n=5$ fermion, or a $n=7$ scalar.
But scalar states may have non-minimal quartic couplings with the Higgs field (see Appendix B). We will then set the 7-plet aside and focus on the fermionic 5-plet for minimality in the following. 

\smallskip

In summary, the `Minimal Dark Matter' construction singles out a 
$$\hbox{fermionic SU(2)$_L$ 5-plet with hypercharge $Y=0$}$$
as providing a fully viable, automatically stable DM particle. It is called `Minimal DM' (MDM) since it is described by the minimal gauge-covariant Lagrangian that one obtains adding the minimal amount of new physics to the Standard Model in order to explain the DM problem.

\section{Cosmological relic density and mass determination}
\label{relic}

Assuming that DM arises as a thermal relic in the Early Universe, via the standard freeze-out process, we can compute the abundance of MDM as a function of its mass $M$. In turn, requiring that MDM makes all the observed DM measured by cosmology, $\Omega_{\rm DM} h^2 =0.110 \pm 0.005$~\cite{cosmoDM}, we can univocally determine $M$. As a general rule of thumb, it is well known~\cite{reviews} that $\Omega_{\rm DM}h^2 \approx 3\cdot 10^{-27} {\rm cm}^3{\rm sec}^{-1}/\langle \sigma_A \beta \rangle$ and that a particle with weak couplings $\alpha_w$ has a $\langle \sigma_A \beta \rangle \approx \alpha_w^2 M_{\rm DM}^{-2}$ that matches $\Omega_{\rm DM}$ for a typical weak scale mass (the so-called WIMP miracle). This is therefore what is to be expected for a pure WIMP model such as MDM is.

\medskip

More precisely, the computation of the relic abundance has to be performed by solving the relevant Boltzmann equation (as we review below) in terms of the detailed annihilation cross section of two MDM particles into any SM state. 
We include the dominant $s$-wave contribution, but also the subdominant $p$-wave, which yields an ${\cal O}$(5\%) correction on the final $\Omega_{\rm DM}$ and the effect of the renormalization of the SM gauge couplings up to the MDM mass scale $M$ (also an ${\cal O}$(5\%) modification). 
On top of this standard computation, however, we have to include the non-perturbative electroweak Sommerfeld corrections, that have a very relevant effect. In fact, this phenomenon significantly enhances non-relativistic annihilations of DM particles with mass $M\circa{>}M_V/\alpha$, when they exchange force mediators of mass $M_V$ with a coupling strenght $\alpha$. In the case of MDM, this is simply the ordinary SM weak force mediated by $W^\pm$ and $Z$, and we will verify `a posteriori' that the relation $M\circa{>}M_{W^\pm}/\alpha_2$ is indeed verified for the MDM masses $M$ that we will obtain. Therefore the Sommerfeld effect is automatically present in the theory and has to be taken into account. We review the basics of the Sommerfeld effect in Appendix A.

\medskip

Let us now describe in more detail the machinery of the computation and the results in our specific case. 
The generic Boltzmann equation that governs the DM abundance as a function of the temperature $T$  reads
 \beq 
 \label{eq:Bol} 
 sZHz \frac{dY}{dz} =
  -2\bigg(\frac{Y^2}{Y^{2}_{\rm eq}}-1\bigg)\gamma_A,\qquad
  \gamma_A =\frac{T}{64 \pi^4} \int_{4M^2 }^{\infty} ds~ s^{1/2}
 {\rm K}_1\bigg(\frac{\sqrt{s}}{T}\bigg) \hat{\sigma}_A(s)
 \eeq
  where $z=M/T$, K$_1$ is a Bessel function, $Z=( 1 - \frac{1}{3} \frac{z}{g_s}\frac{d g_s}{dz})^{-1}$,
  the entropy density of SM particles is $s={2\pi^2}g_{*s}T^3/45$,
  $Y=n_{\rm DM}/s$ where $n_{\rm DM}$ is the number density of DM particles plus anti-particles,
  and $Y_{\rm eq}$ is the value that $Y$ would have in thermal equilibrium.
  The adimensional `reduced annihilation cross section' is defined as 
\beq 
\hat\sigma_A(s)= \int_{-s}^0 dt
\sum \frac{|\mathscr{A}|^2}{8\pi s}
\eeq
where $s,t$ are the Madelstam variables and the sum runs over
all DM components and over all the annihilation channels into all SM vectors, fermions and scalars,
assuming that SM masses are negligibly small.
In the case of MDM, we can write a single equation for the total DM density, in particular neglecting the small splitting $\Delta M$ that we will discuss in Sec.\ref{splitting}, because DM scatterings with SM particles maintain thermal equilibrium within and between the single components.  In this way the formula automatically takes into account all co-annihilations among the multiplet components.
  We also ignore the Bose-Einstein and Fermi-Dirac factors as they are negligible at the temperature $T \sim M/26$ relevant for DM freeze-out.

In full generality, for fermionic DM with $SU(2)_L \times U(1)_Y$ quantum numbers $n$ and $Y$ we get
\begin{eqnarray}
\hat\sigma_A &=& \frac{g_{\cal X}}{24 \pi  n}\nonumber
 \left[(9C_2-21C_1)\beta+(11C_1-5C_2)\beta^3-3 \bigg(2 C_1 (\beta^2-2)+C_2(\beta^2-1)^2\bigg) \ln\frac{1+\beta}{1-\beta }\right]\\
 &&+ g_{\cal X} \bigg(\frac{3g_2^4(n^2-1)+20 g_Y^4 Y^2}{16\pi}+\frac{g_2^4 (n^2-1)+4g_Y^4 Y^2}{128\pi}\bigg)
\bigg(\beta - \frac{\beta^3}{3}\bigg)
\end{eqnarray}
while for scalar DM we get
\begin{eqnarray}
\hat\sigma_A &=&  \frac{g_{\cal X}}{24 \pi  n}\nonumber
 \left[(15 C_1 - 3 C_2) \beta + (5C_2 - 11 C_1) \beta^3+3(\beta^2-1)\bigg(2 C_1+C_2(\beta^2-1)
 \bigg) \ln\frac{1+\beta}{1-\beta }\right]\\
&&+ g_{\cal X} \bigg(\frac{3g_2^4 (n^2-1)+20 g_Y^4 Y^2}{48 \pi}+\frac{g_2^4 (n^2-1)+4g_Y^4 Y^2}{384\pi}\bigg)\cdot \beta^3
\label{eq:sigAs}
\end{eqnarray}
where $x=s/M^2$ and $\beta = \sqrt{1-4/x}$ is defined in the DM DM center-of-mass frame.
The first line gives the contribution of annihilation into vectors,
the second line contains the sum of the contributions of annihilations into
SM fermions and vectors respectively.
The gauge group factors are defined as
\begin{eqnarray}
C_1 &=&\sum_{A,B}  {\rm Tr}\,  T^A T^A T^B T^B =
g_Y^4 nY^4 + g_2^2 g_Y^2 Y^2 \frac{n(n^2-1)}{2}+g_2^4 \frac{n(n^2-1)^2}{16}\\
C_2 &=& \sum_{A,B}  {\rm Tr}\, T^A T^B T^A T^B =  \noindent
g_Y^4 nY^4 + g_2^2 g_Y^2 Y^2 \frac{n(n^2-1)}{2}+g_2^4 \frac{n(n^2-1)(n^2-5)}{16}
\end{eqnarray}
where the sum is over all SM vectors $A=\{Y,W^1,W^2,W^3\}$ with gauge coupling
generators $T^A$.
We have defined $g_{\cal X}$ as the number of DM degrees of freedom for a multiplet with $Y=0$:
$g_{\cal X}=n$ for scalar DM, $g_{\cal X}=2n\, (4n)$ for fermionic Majorana (Dirac) DM.

The DM freeze-out abundance is accurately determined by the leading
two terms of the expansion for small $\beta$, that describe
the $s$-wave and the $p$-wave contributions.
This approximation allows to analytically do the thermal average in eq.\eq{Bol}:
\beq
\hat\sigma_A \stackrel{\beta\to 0}{\simeq} c_s \beta + c_p \beta^3 +\cdots\qquad\hbox{implies}\qquad
 \gamma_A \stackrel{\beta\to 0}{\simeq} 
\frac{MT^3e^{-2M/T}}{32\pi^3}\left[c_s + \frac{3T}{2M}(c_p+\frac{c_s}{2})+\cdots\right].
\eeq

This general formalism can now be easily reduced to the case of the MDM particle, the fermionic 5-plet with $Y=0$. The $s$-wave and $p$-wave coefficients are simply given by
\beq
\label{eq:cs}
c_s = \frac{1035}{8 \pi} g_2^4, \quad c_p = \frac{1215}{8 \pi} g_2^4.
\eeq
The large figures at the numerators simply reflect the `large' number ($n=5$) of components in the multiplet: their coannihilations make for an overall large cross section parameter.

\medskip

\begin{figure}[t]
$$
\includegraphics[width=0.37\textwidth]{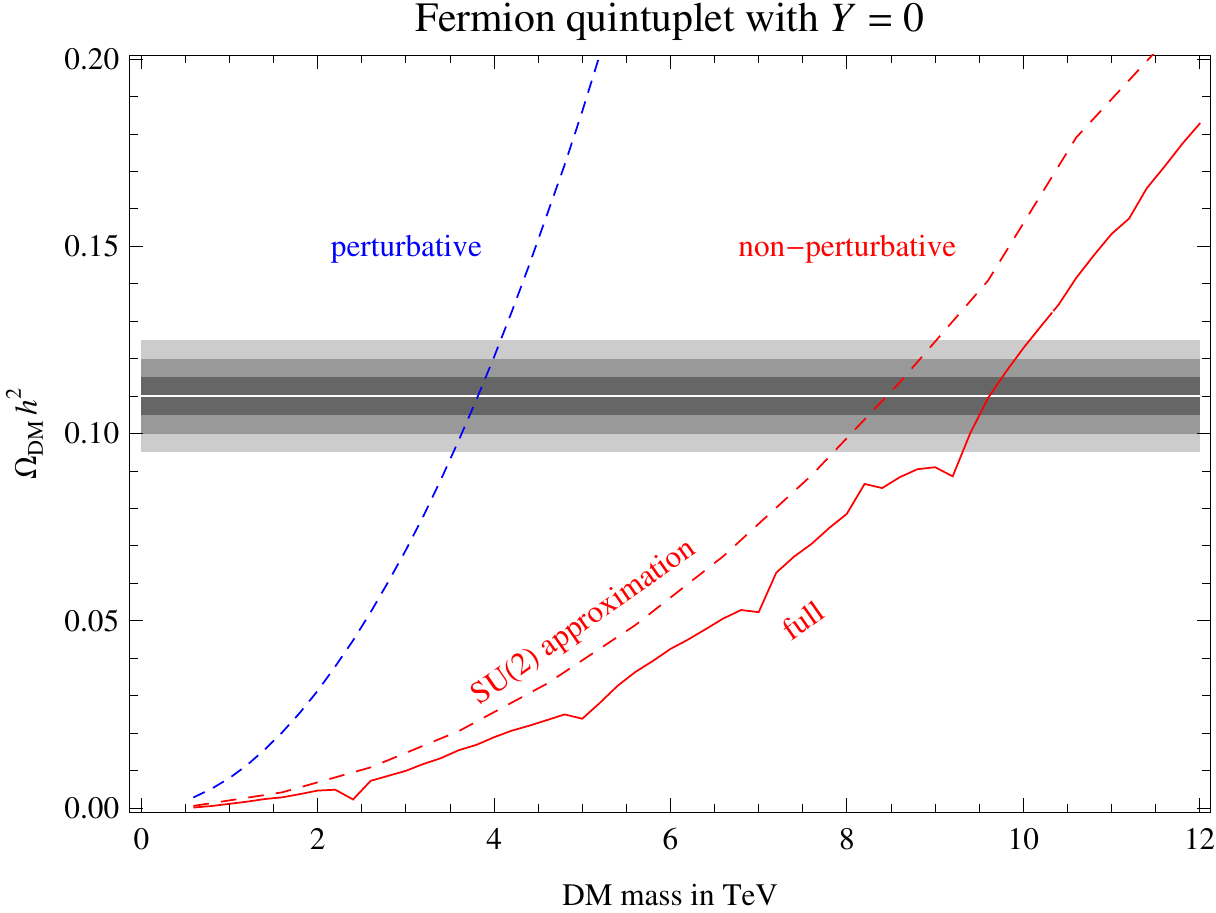}
\includegraphics[width=0.26\textwidth]{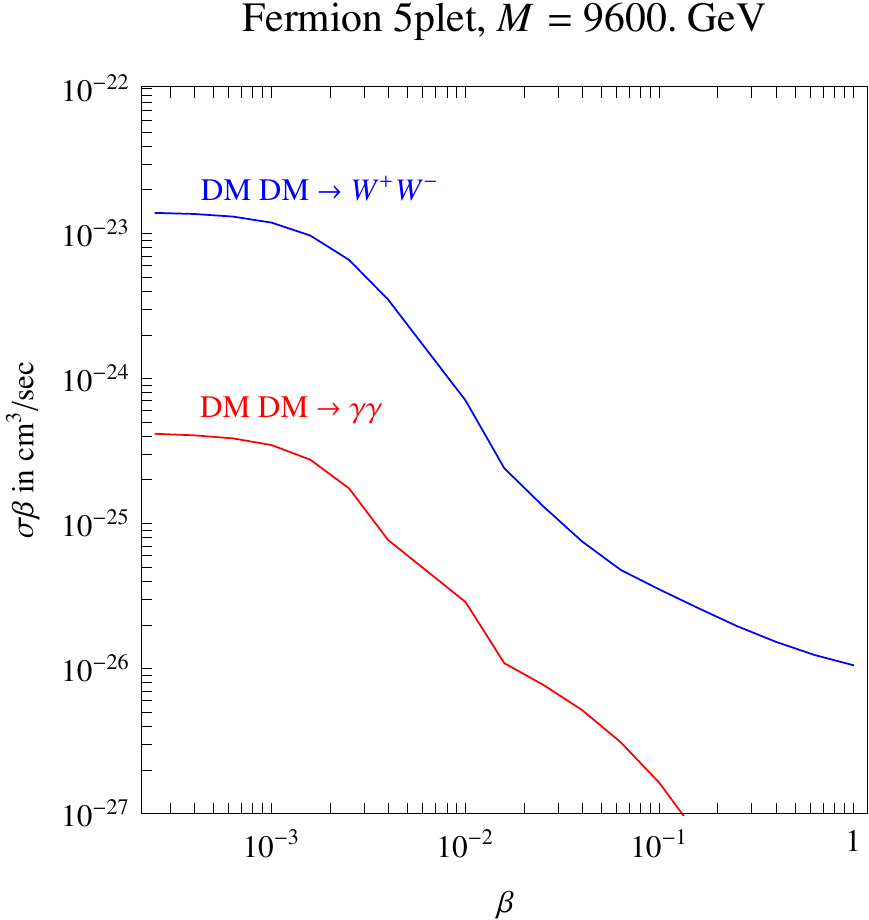}
\includegraphics[width=0.37\textwidth]{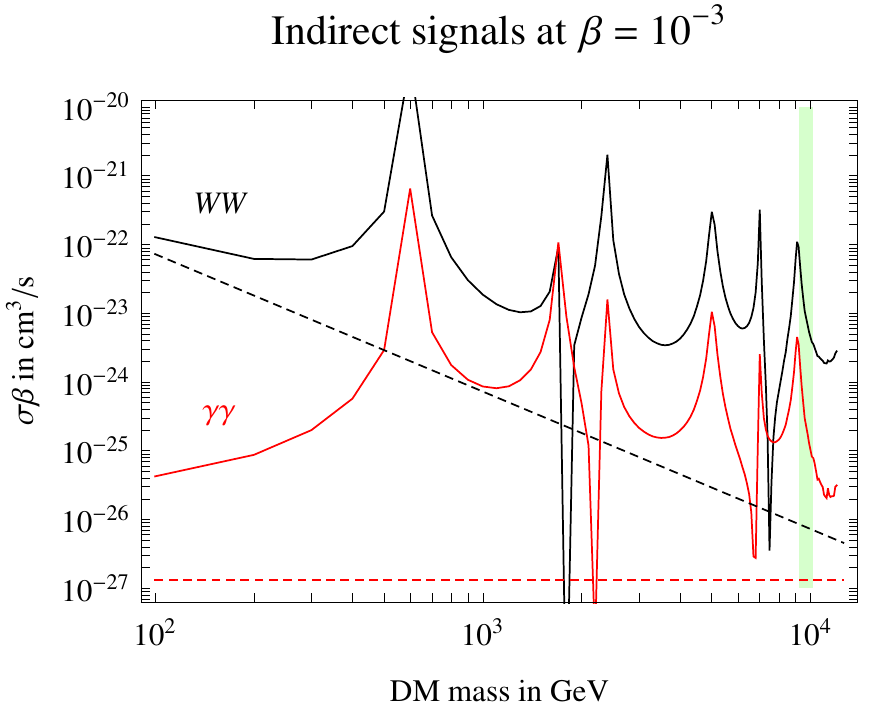}
$$
\caption{\label{fig:relic}\em {\bf Left panel}: The relic abundance of the MDM 5-plet as a function of its mass $M$. The solid red line corresponds to the full result that includes non-perturbative Sommerfeld enhanced annihilations (the dashed red line corresponds to the computation in the {\rm SU(2)}$_L$-symmetric limit), while the dashed blue includes perturbative $s$-wave and $p$-wave tree level annihilations. The horizontal band corresponds to the measured $\Omega_{\rm DM}h^2$, so that its intersection with the red line individuates the DM mass (slightly below 10 TeV) and indicates its uncertainty interval.  
{\bf Central panel}: Velocity dependence of the Sommerfeld-enhanced MDM annihilation cross section.
{\bf Right panel}: The DM\ DM annihilation cross section in the galactic halo today, relevant for indirect detection signatures discussed in Sec.\ref{indirect}. The dashed lines show what the perturbative result would be without Sommerfeld corrections.}
\end{figure}

As anticipated, on top of this the annihilation cross sections are enhanced by the non-perturbative Sommerfeld effect. The effect is strongly dependent on the velocity $\beta$ (in units of $c$) of the DM particles: the total annihilation cross section $\sigma \beta$ grows as $\beta \to 0$ as illustratated in fig.\fig{relic}b. MDM annihilations at the epoch of freeze-out (when $\beta \sim 0.2$) are enhanced by a factor of a few; astrophysical signals due to MDM annihilations in the galaxy today (where $\beta \sim 10^{-3}$) will be much more enhanced, up to a few orders of magnitude, as discussed in Sec.\ref{indirect}.

Solving the Boltzmann equation with all these ingredients allows to compute the relic abundance of the MDM particle $\Omega_{\rm DM} = Y(z \to \infty) s M/\rho_{\rm crit}$ as a function of its mass $M$. The numerical result is plotted in fig.\fig{relic}a. For a given value of the mass, the impact of non-perturbative corrections is relevant, as they enhance the annihilation cross section and therefore reduce the corresponding relic abundance. Matching the relic abundance to the measured $\Omega_{\rm DM} h^2 =0.110 \pm 0.005$ allows therefore to determine  
\beq
M = (9.6 \pm 0.2)\ {\rm TeV}.
\label{eq:M_MDM}
\eeq
For comparison, the value without Sommerfeld corrections (blue dashed curve in fig.\fig{relic}a) would be about 4 TeV.

\bigskip

In summary: the only free parameter of the theory, the DM mass $M$, is fixed by matching the observed relic DM abundance. Not surprisingly, its value turns out to be broadly in the TeV range, because MDM is a pure WIMP model for which the so called `WIMP miracle' applies. The value actually ends up being somewhat higher (eq.\eq{M_MDM}) because the 5-plet has many components so that coannihilations are important {\em and} because Sommerfeld corrections enhance the annihilation cross section.

\subsection{Mass splitting}
\label{splitting}
Supersymmetric models typically feature a model-dependent electroweak-scale mass difference between the neutralino DM candidate and its chargino partners: the (typically multi-GeV) mass difference originates through tree-level mass mixings.
In the MDM case, instead, at tree level all the components of the multiplet have the same mass $M$ computed in the previous section, and then one-loop electroweak corrections make the charged components slightly heavier than the neutral one (if the contrary were true, one would have a charged lightest stable particle, which is of course not phenomenologically allowed).

In full generality, for a fermionic or scalar candidate with overall mass $M$ and hypercharge $Y$, the mass difference induced by loops of SM gauge bosons between two components of ${\cal X}$ with electric charges $Q$ and $Q'$ is found to be
\beq  
M_Q -  M_{Q'} =\frac{\alpha_2 M}{4\pi}\left\{(Q^2-Q^{\prime 2})s_{\rm W}^2 f(\frac{M_Z}{M})+(Q-Q')(Q+Q'-2Y)
\bigg[f(\frac{M_W}{M})-f(\frac{M_Z}{M})\bigg]\right\}
\eeq
where
\beq
f(r) =\left\{\begin{array}{ll}
+r  \left[2 r^3\ln r -2 r+(r^2-4)^{1/2} (r^2+2) \ln A\right]/2 & \hbox{for a fermion}\\
-r \left[2r^3\ln r-k  r+(r^2-4)^{3/2} \ln A\right] /4 & \hbox{for a scalar}
  \end{array}\right.
  \eeq   
  with $A = (r^2 -2 - r\sqrt{r^2-4})/2$ and $s_{\rm W}$ the sine of the weak angle.
In the numerically relevant limit $M\gg M_{W,Z}$ the one-loop corrections get the universal value $f(r)\stackrel{r\to 0}{\simeq}  2\pi r$. For the fermionic 5-plet with $Y=0$, the splitting between the neutral component (the DM particle) and its $Q=\pm 1$ partners equals therefore to
\beq 
\Delta M = \alpha_2 M_W \sin^2\frac{\theta_{\rm W}}{2}=(166\pm 1)\MeV.
\label{eq:166}
\eeq

In general, the mass splitting between charged and neutral components can be intuitively understood in terms of the {\em classical} non-abelian Coulomb energy (the energy stored in the electroweak electric fields that a point-like charge at rest generates around itself, which can be thought as an additional contribution to its mass with respect to the mass of an equivalent neutral particle). 
Indeed, for a scalar or fermion with a gauge coupling $g$ under a vector with mass $M_V$, the Coulomb energy is:
$$  \delta M = \int d^3r \left[\frac{1}{2}(\vec\nabla \varphi)^2 + \frac{M_V}{2}\varphi^2\right]=
\frac{\alpha}{2} M_V + \infty\qquad
\varphi(r) = \frac{g e^{-M_V r/\hbar}}{4\pi r}$$
As SU(2)$_L$ invariance is restored at distances $r\ll 1/M_{W,Z}$, the UV divergent term cancels out when 
computing the correction to the intra-multiplet mass splitting.
Therefore we understand why the effect in the limit $M\gg M_{W,Z}$ does not depend on the DM spin, and why the
neutral component is lighter than the charged ones.
This intuitive picture allows us to guess that, when considering the low-energy DM/nuclei relevant for DM direct detection,
the spin-independent cross section will be suppressed only by $1/M_{W,Z}^2$ (and not by the much smaller $1/M^2$),
because what scatters is the cloud of electro-weak electric fields that extend up to $r\sim1/M_{W,Z}$.
This agrees with the Feynman diagram computation presented in the next section.

\section{Direct Detection signatures}
\label{direct}

\begin{figure}[t]
$$
\includegraphics[width=\textwidth]{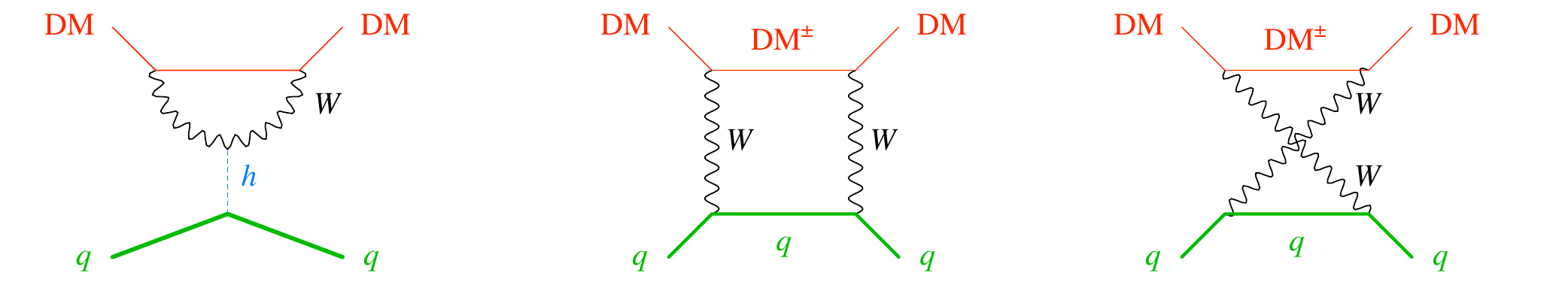}$$
\caption{\label{fig:FeynLoop}\em One loop DM/quark scattering for fermionic MDM with $Y=0$ (two extra graphs involving the four particle vertex exist in the case of scalar MDM).}
\end{figure}

Direct searches for DM aim to detect the recoils of nuclei in a low background detector produced by the rare collisions of DM halo particles on such nuclei. So far, the DAMA/Libra experiment~\cite{DAMA} has reported the detection of an annual modulation of the total number of events compatible with the effect that the motion of the Earth in the DM halo would produce. The CDMS~\cite{CDMS} and Xenon~\cite{Xenon} experiments have published exclusion bounds.

\bigskip

As discussed in Sec.\ref{model}, MDM candidates with $Y=0$ have vanishing $\DM \,{\cal N}$ direct detection cross sections at tree level (see eq.\eq{direct}). The scattering on nuclei ${\cal N}$ proceeds therefore at one-loop, via the diagrams in fig.\fig{FeynLoop} that involve one of the charged components ${\cal X}^\pm$ of the multiplets.
An explicit  computation of these one-loop diagrams is needed
to understand qualitatively and quantitatively the resulting cross section.
Non-relativistic MDM/quark interactions
of fermionic ${\cal X}$ with mass $M\gg M_W \gg m_q$
are described by the effective on-shell Lagrangian
\beq
\label{eq:directW} \Lag_{\rm eff}^W =(n^2-(1\pm 2Y)^2)\frac{\pi \alpha_2^2}{16 M_W}\sum_q
\left[
\left(\frac{1}{M_W^2}+\frac{1}{m_h^2}\right)  [\bar {\cal X}{\cal X}] m_q [\bar{q}{q}]-\frac{2}{3M}
{ [\bar {\cal X}\gamma_\mu\gamma_5 {\cal X}][\bar {q}\gamma_\mu\gamma_5 {q}] }\right]
\eeq
where the $+$ ($-$) sign holds for down-type (up-type) quarks $q=\{u,d,s,c,b,t\}$, 
$m_h $ is the Higgs mass and $m_q$ are the quark masses.
The first operator gives dominant spin-independent effects and is not suppressed by $M$;
the second operator is suppressed by one power of $M$ and gives spin-dependent effects.
Parameterizing the nucleonic matrix element as
\beq 
\langle{N} | \sum_q m_q \bar{q}q | N\rangle\equiv  f m_N  \eeq
where $m_N$ is the nucleon mass, the spin-independent DM cross section on a target nucleus ${\cal N}$ with mass $M_{\cal N}$
is given by
\beq\label{eq:sigmaSI} 
\sigma_{\rm SI}(\DM\,{\cal N}\to \DM\,{\cal N})
=(n^2-1)^2 \frac{\pi\alpha_2^4 M_{\cal N}^4   f^2}{64 M_W^2}\left(\frac{1}{M_W^2}+\frac{1}{m_h^2}\right)^2.
\label{eq:direct}
\eeq
\begin{figure}[t]
$$\hspace{-5mm}
\includegraphics[width=0.8\textwidth]{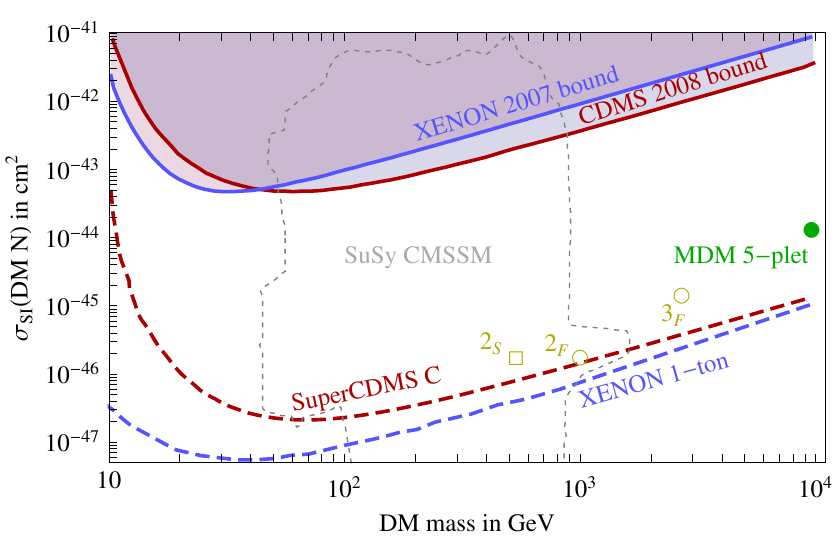}$$
\caption{\label{fig:direct}\em
{\bf Parameter space of the spin-independent cross section per nucleon}: the MDM 5-plet is represented by the green dot. The predictions is univocal (having assumed that the matrix element $f=1/3$ and $115$ GeV as a value for the higgs mass), while typical minimal SUSY models span a large portion of the parameter space, suggested here by the area with dotted contours.
The dashed lines indicate the sensitivity of some future experiments~\cite{future}. Other non-minimal candidates, discussed in Appendix B, are reported for completeness: the fermionic 3-plet and the fermion and scalar doublet.}
\end{figure}
The case of scalar ${\cal X}$  is not much different: 
the $M$-independent contribution to $\sigma_{\rm SI}$ is equal to
the fermionic result of eq.\eq{sigmaSI} but there is no spin-dependent effect.

Assuming $m_h=115\GeV$ and
$f\approx 1/3$
(QCD uncertainties induce a one order of magnitude indetermination on $\sigma_{\rm SI}$\footnote{More precisely, one needs to
consider the effective Lagrangian for off-shell quarks, finding various operators that become equivalent only on-shell.
Their nucleon matrix elements can differ; we ignore this issue because presently it is within the QCD errors.})
we find therefore for the fermionic MDM 5-plet 
\beq
\sigma_{\rm SI} = 1.2 \cdot 10^{-44}\ {\rm cm}^2.
\eeq
As usual~\cite{reviews,Drees,Pittel}, $\sigma_{\rm SI}$ is defined to be the cross section per nucleon. 
The prediction is a definite number (as opposed to the large areas in the plane $M$/$\sigma$ that is covered by typical supersymmetric constuctions by varying the model parameters) and Fig.\fig{direct} shows that this value is within or very close to the sensitivities of experiments currently under study, such as Super-CDMS and Xenon 1-ton~\cite{future}. 
The annual modulation effect of the DAMA/Libra experiment~\cite{DAMA} cannot be explained by MDM candidates, since they have too large masses and too small cross sections with respect to the properties of a WIMP compatible with the effect.

\section{Indirect Detection signatures}
\label{indirect}

Indirect searches are one of the most promising ways to detect Dark Matter. DM particles in the galactic halo are expected to annihilate and produce fluxes of cosmic rays that propagate through the galaxy and reach the Earth.
Their energy spectra carry important information on the nature of the DM particle (mass and primary annihilation channels). Many experiments are searching for signatures of DM annihilations in the fluxes of $\gamma$ rays, positrons and antiprotons and there has been recently a flurry of experimental results in this respect:
\begin{itemize}
\item data from the PAMELA satellite show a steep increase in the energy spectrum of the positron fraction $e^+/(e^++e^-)$ above  10~GeV up to 100~GeV~\cite{PAMELApositrons}, compatibly with previous less certain hints from HEAT~\cite{HEAT} and AMS-01~\cite{AMS01};
\item data from the PAMELA also show no excess in the $\bar p/p$ energy spectrum~\cite{PAMELApbar} compared with the predicted background;
\item the balloon experiments ATIC-2~\cite{ATIC-2} and PPB-BETS~\cite{Torii:2008xu} report the presence of a peak in the $e^++e^-$ energy spectrum at around  500-800~GeV;
\item the HESS telescope has also reported the measurement~\cite{HESSleptons} of the $e^++e^-$ energy spectrum above energies of 600 GeV up to a few TeV: the data points show a steepening of the spectrum which is compatible both with the ATIC peak (which cannot however be fully tested) and with a power law with index $-3.05 \pm 0.02$ and a cutoff at $\approx$2 TeV. 
\end{itemize}

Some words of caution apply to the balloon results: the Monte Carlo simulations that such experiments need in order to tag $e^\pm$ and infer their energy have been tested only up to LEP energies; the excess is based on just a few data-points that are not cleanly consistent between ATIC-2 and the smaller PPB-BETS; emulsion chambers (EC) balloon experiments~\cite{EC} do not show evidence for an excess, although they have larger uncertainties. The upcoming results of the FERMI/LAT mission on the $e^++e^-$ energy spectrum~\cite{FERMIleptons} will hopefully soon indicate in a more definitive and precise way the shape of the spectrum.\footnote{{\bf Note added.} See footnote~\ref{footnoteFermi}.}

In full generality~\cite{CKRS}, the PAMELA positron and anti-proton data, if interpreted in terms of DM, indicate either (i) a DM particle of any mass (above about 100 GeV) that annihilates only into leptons, not producing therefore unseen antiprotons or (ii) a DM particle with a mass around or above 10 TeV, that can annihilate into any channel. Adding the balloon data, the mass is pinned down to about 1 TeV and only leptonic channels are allowed.

\medskip

Minimal Dark Matter has definite predictions for the fluxes of positrons, antiprotons and gamma rays, presented in~\cite{MDM3} before the announcements of the experimental results (and even before some of the relevant experiments, such as FERMI, started taking data). The comparison with the data that are now available allows to fix the astrophysical uncertainties and test the model.
The MDM 5-plet annihilates at tree level into $W^+W^-$, and at loop level into
$\gamma\gamma$, $\gamma Z$, $ZZ$.
The relative cross-sections are significantly affected by the non-perturbative Sommerfeld corrections discussed in Sec.\ref{relic} and in Appendix A. The best-fit values are: 
\beq
\label{eq:annihilationsigma}
\langle \sigma v \rangle_{WW} = 1.1\cdot 10^{-23} {{\rm cm}^3 \over {\rm sec}},\qquad
\langle \sigma v \rangle_{\gamma\gamma}= 3\cdot 10^{-25} {{\rm cm}^3\over {\rm sec}}
\eeq
Annihilation cross sections into $\gamma Z$ and $ZZ$ are given by
\beq \sigma_{\gamma Z} = 2\sigma_{\gamma\gamma}/\tan^2\theta_{\rm W}=
6.5 \sigma_{\gamma\gamma}
,\qquad
\sigma_{ZZ} = \sigma_{\gamma\gamma}/\tan^4\theta_{\rm W} = 10.8\sigma_{\gamma\gamma}
\eeq
(that actually hold for any MDM candidate with $Y=0$).
However, as shown in fig.\fig{relic}c, as a consequence of Sommerfeld corrections the DM DM annihilation cross sections exhibit a quite steep dependence on $M_\DM$ and can vary by about one order of magnitude around these central values within the range allowed at $3\sigma$ by the cosmological DM abundance. The cross section also depends on the DM velocity $v$, reaching a maximal value for $v\to 0$, as shown in fig.\fig{relic}b. The average DM velocity in our galaxy, $v\approx 10^{-3}$, is however low enough that $\sigma v$ is close to its maximal value, which we assume.

\begin{figure}[t]
\begin{center}
\includegraphics[width=\textwidth]{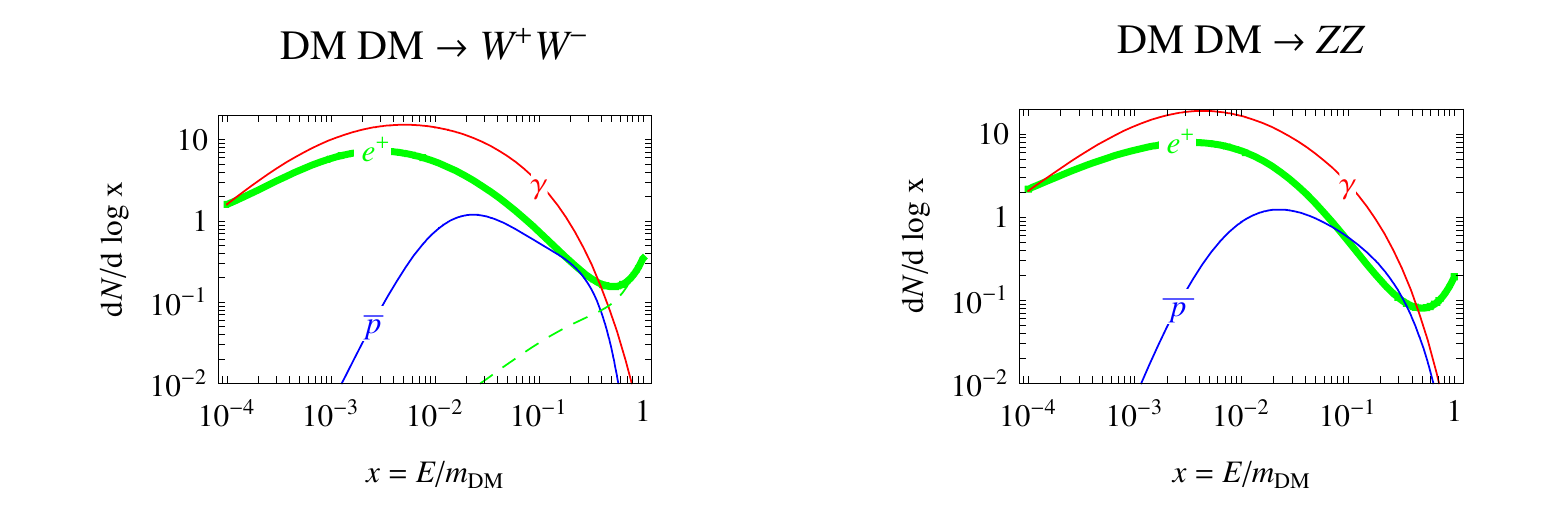}
\caption{\em\label{fig:Fragmentation} Energy spectra of $e^+,\bar p, \gamma$ produced by non-relativistic
$\DM\, \DM$ annihilations into $W^+W^-$ and into $ZZ$, the only two relevant channels for MDM (together with $\gamma\gamma$, which produces a $\gamma$ line at the MDM mass, and $\gamma Z$, which can be deduced from the $\gamma\gamma$ and $ZZ$ channels).
The $e^+$ have a secondary component (dashed green line shown on the $W^+W^-$ plot), that dominates at large $x\sim 1$.
}
\end{center}
\end{figure}

The energy spectra of $e^+$, $\bar p$ and $\gamma$ per annihilation at production, as generated by PYTHIA supplemented by appropriate custom routines that allow to take into account the spin correlations of SM vectors\,\footnote{More details are given in ref.~\cite{MDM3}.}, are represented in fig.\fig{Fragmentation}. We next need to consider where these fluxes of particles are produced in the galaxy and how they propagate to the Earth.

\subsection{Positrons}

\begin{figure}[t]
\begin{center}
\includegraphics[width=0.45\textwidth]{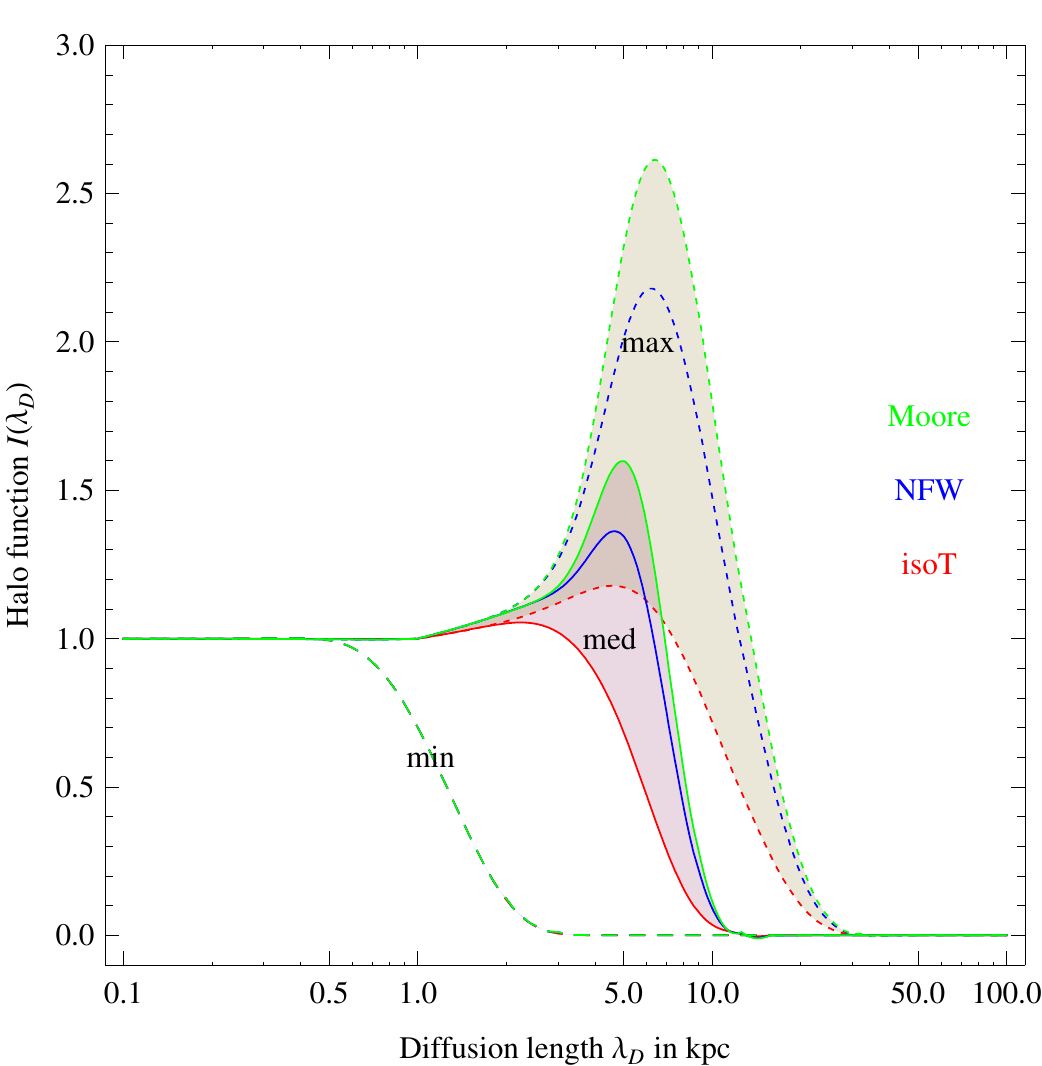}\qquad
\includegraphics[width=0.45\textwidth]{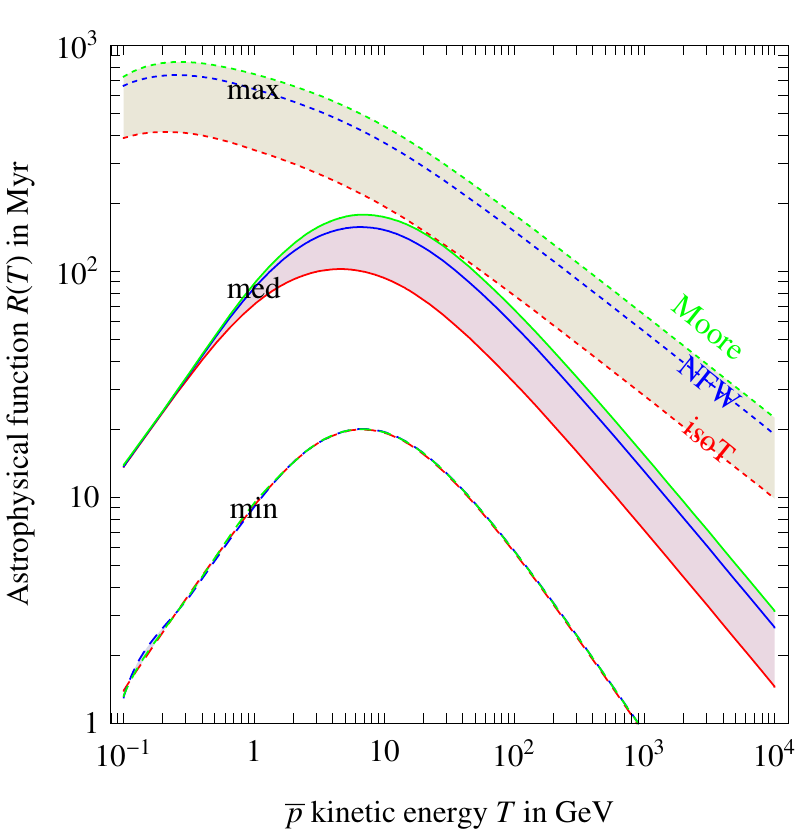}
\caption{\em\label{fig:HaloI} {\bf The astrophysical propagation functions for positrons and antiprotons.} Left: The `halo function' $I(\lambda_D)$ of eq.\eq{fluxpositrons} that encodes the astrophysics of DM DM annihilations into positrons and their propagation up to the Earth. The diffusion length is related to energy losses as in eq. (14) of~\cite{MDM3}, that also provides fit functions for all cases.
Right: The somewhat analogous $\bar p$ astrophysical function $R(T)$ of eq.\eq{RT}. In both cases, the dashed (solid) [dotted] bands assumes the {\rm min (med) [max]} propagation configuration of table \ref{tab:proparam}.
Each  band contains 3 lines, that correspond to the isothermal (red lower lines), NFW (blue middle lines) and Moore (green upper lines) DM density profiles.
}
\label{default}
\end{center}
\end{figure}

The positron flux per unit energy from DM annihilations in any point in space and time is given by $\Phi_{e^+}(t,\vec x,E) = v_{e^+} f/4\pi$ (units $1/\GeV\cdot{\rm cm}^2\cdot{\rm s}\cdot{\rm sr}$)
where $v_{e^+}$ is the positron velocity (essentially equal to $c$ in the regimes of our interest) and the positron number density per unit energy, $f(t,\vec x,E)= dN_{e^+}/dE$,
obeys the diffusion-loss equation:
\beq \label{eq:diffeq}\frac{\partial f}{\partial t}-
K(E)\cdot \nabla^2f - \frac{\partial}{\partial E}\left( b(E) f \right) = Q\eeq
with diffusion coefficient $K(E)=K_0 (E/\GeV)^\delta$
and energy loss coefficient $b(E)=E^2/(\GeV\cdot \tau_E)$ with $\tau_E = 10^{16}\,{\rm s}$.
They respectively describe transport through the turbulent magnetic fields and energy loss due to
synchrotron radiation and inverse Compton scattering on CMB photons and on infrared galactic starlight.
Eq.~(\ref{eq:diffeq}) is solved in a diffusive region with the shape of a solid flat cylinder that sandwiches the galactic plane, with height $2L$ in the $z$ direction and radius $R=20\,{\rm kpc}$ in the $r$ direction~\cite{DiffusionCylinder}. The location of the solar system corresponds to $\vec x  = (r_{\odot}, z_{\odot}) = (8.5\, {\rm kpc}, 0)$.
The boundary conditions impose that the positron density $f$ vanishes on the surface of the cylinder, outside of which positrons freely propagate and escape.
Values of the propagation parameters $\delta$, $K_0$ and $L$ are deduced from a variety of cosmic ray data and modelizations. They represent a source of uncertainty over which to scan in order to reach the final predictions for the fluxes.
We consider the sets presented in Table~\ref{tab:proparam}~\cite{FornengoDec2007}.
\begin{table}[t]
\center
\begin{tabular}{c|cc|ccc|c}
 & \multicolumn{2}{c|}{Positrons} & \multicolumn{3}{c|}{Antiprotons}  \\
Model  & $\delta$ & $K_0$ (kpc$^2$/Myr) & $\delta$ & $K_0$ (kpc$^2$/Myr) & $V_{\rm conv}$ (km/s) & $L$ (kpc)  \\
\hline 
MIN  & 0.55 & 0.00595 & 0.85 &  0.0016 & 13.5 & 1 \\
MED & 0.70 & 0.0112 & 0.70 &  0.0112 & 12 & 4  \\
MAX  & 0.46 & 0.0765 &  0.46 &  0.0765 & 5 & 15 
\end{tabular}
\caption{\em {\bf Propagation parameters} for charged (anti)particles in the galaxy (from~\cite{FornengoDec2007}, \cite{DonatoPRD69}). 
\label{tab:proparam}}
\end{table}
Finally, the source term due to DM DM annihilations in each point of the halo with DM density $\rho(\vec x)$ is
\beq \label{eq:Q}
Q = \frac{1}{2} \left(\frac{\rho}{M_{\rm DM}}\right)^2 f_{\rm inj},\qquad f_{\rm inj} = \sum_{k} \langle \sigma v\rangle_k \frac{dN_{e^+}^k}{dE}\eeq
where $k$ runs over all the channels with positrons in the final state, with the respective thermal averaged cross sections $\sigma v$. Several galactic DM profiles are computed on the basis of numerical simulations: isothermal~\cite{isoT}, Einasto~\cite{Einasto}, Navarro-Frenk-White~\cite{NFW}, Moore~\cite{Moore} (roughly in order of cuspiness at the galactic center). The choice of profile introduces a further element of astrophysical uncertainty over which to scan.\\
The solution for the positron flux at Earth can be written in a useful semi-analytical form~\cite{FornengoDec2007,HisanoAntiparticles}:
\beq 
\Phi_{e^+}(E,\vec r_{\odot}) = B \frac{v_{e^+}}{4\pi b(E)}
\frac{1}{2} \left(\frac{\rho_\odot}{M_{\rm DM}}\right)^2  \int_{E}^{M_{\rm DM}}  dE'~f_{\rm inj}(E')\cdot  I \left(\lambda_D(E,E')\right)
\label{eq:fluxpositrons}
\eeq
where $B\ge 1$ is an overall boost factor discussed below, $\lambda_D(E,E')$ is the diffusion length from energy $E'$ to energy $E$.
The adimensional `halo function' $I(\lambda_D)$~\cite{FornengoDec2007} fully encodes the galactic astrophysics and is independent on the particle physics model. Its possible shapes are plotted in fig.\fig{HaloI}a for most common choices of DM density profiles and set of positron propagation parameters.

\smallskip

The flux of positrons from DM annihilations has to be summed to the expected astrophysical background. We take the latter from the CR simulations of~\cite{MoskalenkoStrong} as parameterized in~\cite{bkgpositrons} by $\Phi_{e^+}^{\rm bkg} = 4.5\, E^{0.7}/(1+650\, E^{2.3}+1500\, E^{4.2})$ for positron and 
$\Phi_{e^-}^{\rm bkg}$ = $\Phi_{e^-}^{\rm bkg,\, prim}$  +$ \Phi_{e^-}^{\rm bkg,\, sec}$ = $0.16\, E^{-1.1}/(1+11\, E^{0.9}+3.2\, E^{2.15})$ + $0.70\, E^{0.7}/(1+110\, E^{1.5}+580\, E^{4.2})$ for electrons, with $E$ always in units of GeV. 
These not-so-recent background computations have recently been revised and questioned~\cite{Delahaye}: background shapes with a downturn around energies of a few GeV have been investigated in order to incorporate the PAMELA excess as a feature of the background. 

Finally, the DM density in our galaxy might have local clumps that would enhance the positron flux by an unknown
`boost factor' $B \ge 1$. We take it as energy independent: this is a simplifying (but widely used) assumption. Detailed recent studies~\cite{Lavalle1,Lavalle2,minispikes,Berez} find that a certain energy dependance can be present, subject to the precise choices of the astrophysical parameters. Within the uncertainty, these studies also converge towards small values of $B$ (except for extreme scenarios), with $B = {\cal O}(10)$ still allowed.

\begin{figure}[t]
$$\hspace{-5mm}
\includegraphics[width=0.5\textwidth]{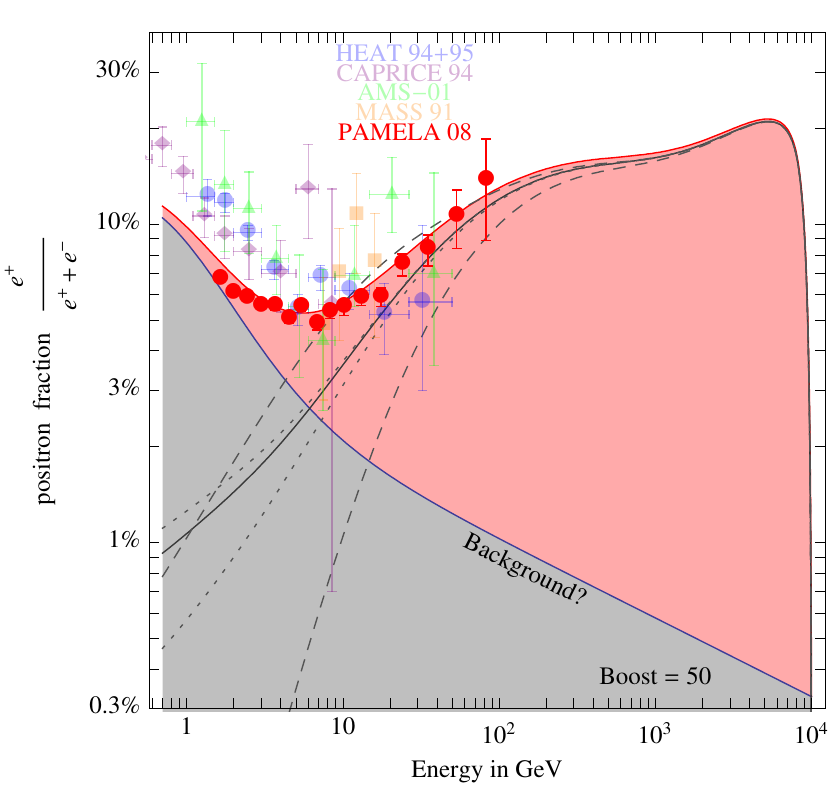}$$
\caption{\label{fig:positrons}\em
{\bf Positron fraction}. The positron fraction from MDM galactic annihilations is compared with the data from PAMELA and previous experiments. We have taken a boost factor of 50 (see text). We show the DM signal (the lines) and the total $e^+$ fraction when summed to the background (shaded area). The main result (solid line and shaded area) is computed assuming a benchmark NFW DM profile and MED propagation parameters. The fainter dashed lines correspond to changing the propagation parameters to MIN (lower) and MAX (upper). The fainter dotted lines correspond to changing the DM profile to isothermal (lower) and Moore (upper).}
\end{figure}

\smallskip

On the basis of the ingredients above, the fluxes at Earth of positrons from MDM annihilations can be compared with the experimental results. This is shown in fig.\fig{positrons}. One sees that the predictions agree very well with the PAMELA results on the whole range of energies. We have assumed $B=50$, which is the value found to provide the best fit to positrons, electrons and antiprotons data (discussed below) combined. This is quite a large value, in tension with the determinations discussed above. On the other hand, lower values (down to about 20) would still give a reasonably good fit and in any case the MDM annihilation cross sections of eq.\eq{annihilationsigma} carry a one order-of-magnitude uncertainty. In order to be conservative, we prefer to quote the boost value `as is', instead of looking for possible optimizations.
The figure also illustrates that the DM signal is only very mildly affected by changing the DM density profile (dotted lines). It somewhat depends on the $e^+$ propagation model in our galaxy (dashed lines); this uncertainty will be reduced by future measurements of cosmic rays and is however present only at $E\ll M_\DM$.

\subsection{Electrons + positrons}

\begin{figure}[t]
$$\hspace{-5mm}
\includegraphics[width=0.5\textwidth]{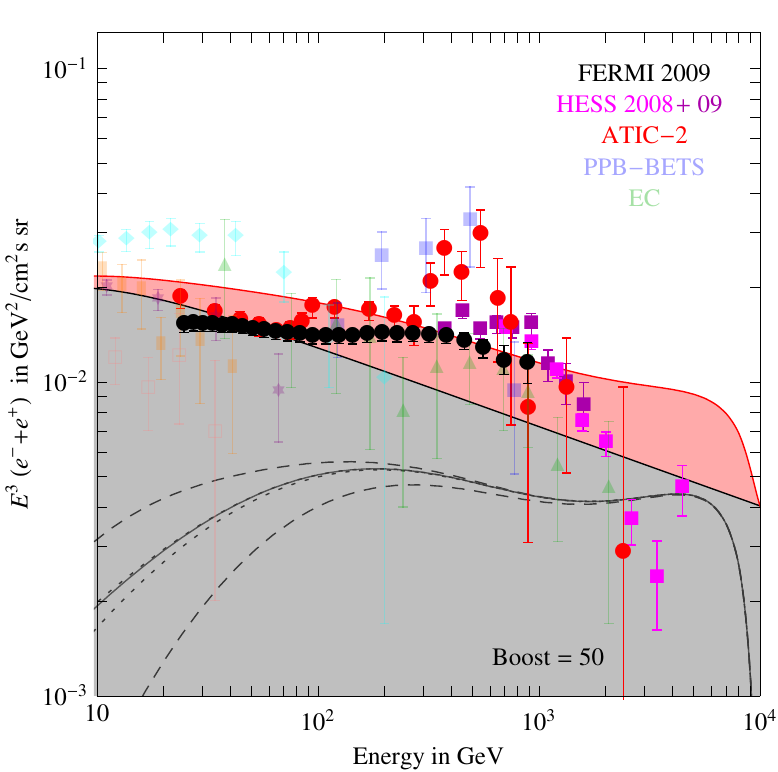}$$
\caption{\label{fig:leptons}\em
{\bf Sum of electrons and positrons}. The flux of electrons and positrons from MDM annihilations is compared to the data from ATIC, PPB-BETS, EC and previous experiments, plus the datapoints from HESS and FERMI. The main result (solid line and shaded area) is computed assuming a benchmark NFW DM profile and MED propagation parameters. The fainter dashed lines correspond to changing the propagation parameters to MIN (lower) and MAX (upper). The fainter dotted lines correspond to changing the DM profile to isothermal (lower) and Moore (upper).}
\end{figure}

The computation of the fluxes of $e^++e^-$ from MDM annihilations is just a rearrangement of the calculations for positrons presented in the previous section. 
Fig.\fig{leptons} shows the predicted flux as compared to the results from the balloon experiments ATIC, PPB-BETS and EC and HESS. It is apparent that the MDM predictions are not compatible with the peak individuated (mainly) by the ATIC data points: a spectrum which is flat up to the higher energies, with a smooth endpoint somewhat below $M$ would be expected. The HESS datapoints indicate a steepening of the spectrum with respect to GeV energies. They are compatible with the shoulder of the ATIC peak, which however cannot be fully tested.\\
If the presence of the ATIC peak will be confirmed in future data sets, therefore strongly indicating a DM mass around 1 TeV, MDM will be falsified. 

Assuming that the astrophysical background is a power-law 
(HESS data however indicate a steepening around 1 TeV), the 
MDM 5-plet predicts a slightly-harder
quasi-power-law $e^++e^-$ spectrum up to several TeV energies.
\footnote{{\bf Note added}.\label{footnoteFermi}
The ATIC peak, that was incompatible with Minimal Dark Matter, is now contradicted by the new more precise FERMI data~\cite{FERMI} (unless an exceptionally bad energy resolution is assumed for FERMI),
that we just superimposed to fig.\fig{leptons}, without modifying the MDM prediction to better fit the new data.
The problem is now that FERMI data are consistent with the steepening apparent in the HESS data (supplemented by the new results at lower energy in~\cite{HESS2009}).
If this feature will be confirmed and cannot be attributed to the astrophysical background, Minimal Dark Matter 
will be excluded as an interpretation of the PAMELA and FERMI $e^\pm$ excesses.
}

\begin{figure}[t]
$$\hspace{-5mm}
\includegraphics[width=0.5\textwidth]{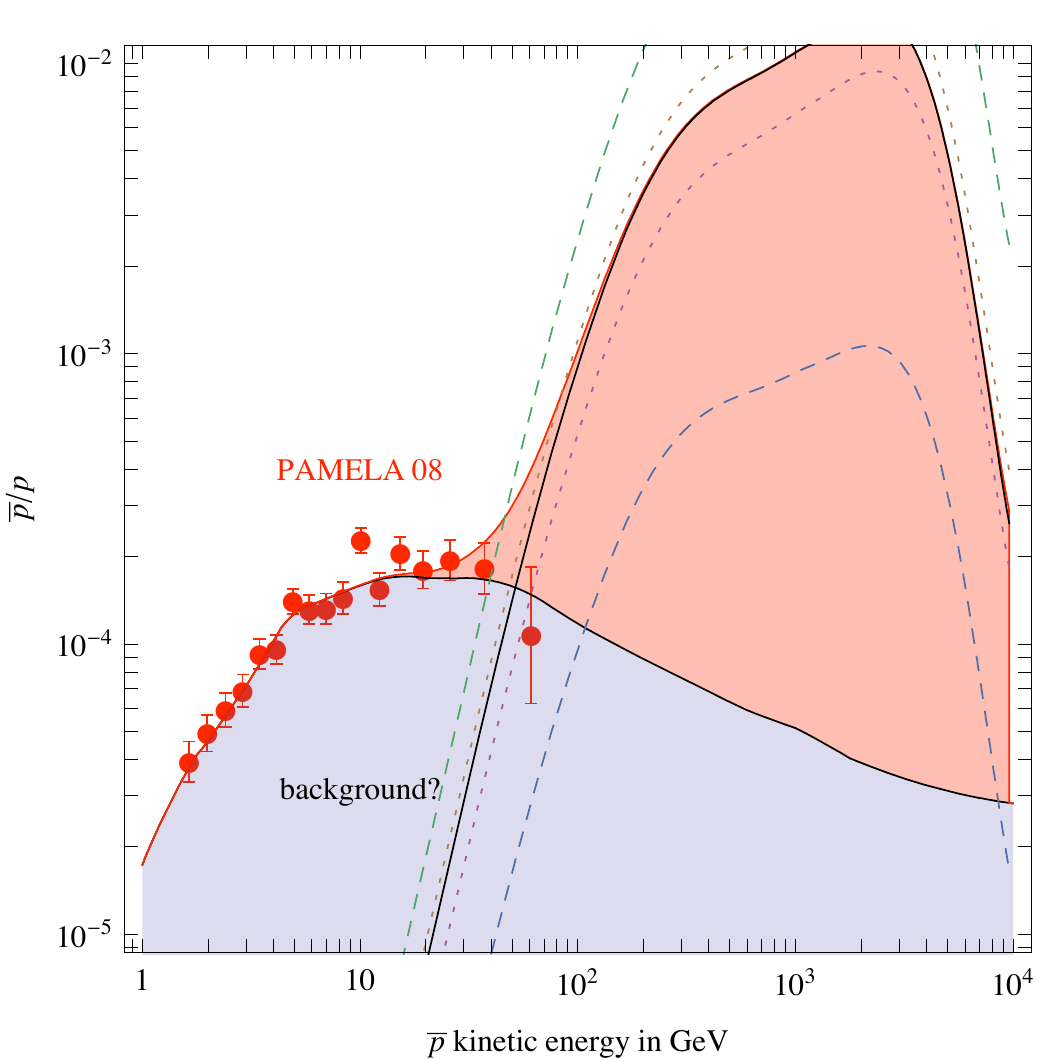}$$
\caption{\label{fig:antiprotons}\em
{\bf Antiprotons}. The antiproton over proton ratio (at top of the atmosphere) from MDM annihilations, compared to the recent PAMELA data. We have assumed the same boost factor as for positrons ($B = 50$). The main result (solid line and shaded area) is computed assuming a benchmark NFW DM profile and MED propagation parameters. The fainter dashed lines correspond to changing the propagation parameters to MIN (lower) and MAX (upper). The fainter dotted lines correspond to changing the DM profile to isothermal (lower) and Moore (upper).}
\end{figure}

\subsection{Antiprotons}
The propagation of anti-protons through the galaxy is described by a diffusion equation analogous to the one for positrons.
Again, the number density of anti-protons per unit energy $f(t,\vec x,T) = dN_{\bar p}/dT$ vanishes on the surface of the cylinder at $z=\pm L$ and $r=R$. $T=E-m_p$ is the $\bar p$ kinetic energy, conveniently used instead of the total energy $E$ (a distinction which will not be particularly relevant for our purposes as we look at energies much larger than the proton mass $m_p$).
Since $m_p\gg m_e$ we can neglect the energy loss term, and the diffusion equation for $f$ is
\beq 
\label{eq:diffeqp}
\frac{\partial f}{\partial t} - K(T)\cdot \nabla^2f + \frac{\partial}{\partial z}\left( {\rm sign}(z)\, f\, V_{\rm conv} \right) = Q-2h\, \delta(z)\, \Gamma_{\rm ann} f . 
\eeq
The pure diffusion term can again be written as $K(T) = K_0 \beta \, (p/\GeV)^\delta$, where $p = (T^2 +2 m_p T)^{1/2}$ and 
$\beta = v_{\bar p}/c = \left(1-m_p^2/(T+m_p)^2\right)^{1/2}$ are the antiproton momentum and velocity.  The $V_{\rm conv}$ term corresponds to a convective wind, assumed to be constant and directed outward from the galactic plane, that tends to push away $\bar p$ with energy $T \circa{<}10\, m_p$. The diferent sets of values of the parameters are given in table~\ref{tab:proparam}.
The last term in eq.\eq{diffeqp} describes the annihilations of $\bar p$ on interstellar protons in the galactic plane
(with a thickness of $h=0.1\,{\rm kpc} \ll L$) with rate
$\Gamma_{\rm ann} = (n_{\rm H} + 4^{2/3} n_{\rm He}) \sigma^{\rm ann}_{p\bar{p}} v_{\bar{p}}$,
where $n_{\rm H}\approx 1/{\rm cm}^3$ is the hydrogen density, $n_{\rm He}\approx 0.07\, n_{\rm H}$ is the Helium density (the factor $4^{2/3}$ accounting for the different geometrical cross section in an effective way)
and the $\sigma^{\rm ann}_{p\bar{p}}$ given esplicitely in ~\cite{MDM3, crosssection,HisanoAntiparticles}.
We neglect the effect of ``tertiary anti-protons''. This refers to primary $\bar p$ after they have undergone non-annihilating interactions on the matter in the galactic disk, losing part of their energy. 

In the ``no-tertiaries" approximation that we adopt, the solution~\cite{methodPbar, TailletRRDA, MaurinApJ555} for the antiproton flux at the position of the Earth $ \Phi_{\bar p}(T,\vec r_\odot) = v_{\bar p}/(4\pi) f $ acquires a simple factorized form (see e.g.~\cite{DonatoPRD69})
\beq
\Phi_{\bar p}(T,\vec r_\odot) = B \frac{v_{\bar p}}{4\pi}  \left(\frac{\rho_\odot}{M_{\rm DM}}\right)^2 R(T)   \sum_k \frac{1}{2} \langle \sigma v\rangle_k \frac{dN^k_{\bar p}}{dT}
\label{eq:RT}
\eeq
where $B$ is the boost factor. The $k$ index runs over all the annihilation channels with anti-protons in the final state, with the respective cross sections; this part contains the particle physics input. The function $R(T)$ encodes all the astrophysics and depends on the choice of halo profile and propagation parameter set. It is plotted in fig.\fig{HaloI} for several possible choices.
Finally, for completeness we also take into account the solar modulation effect, due to the interactions with the solar wind, that distorts the spectrum via a slight increase of the low energy tail, as described in more detail in~\cite{GA,MDM3}.

The astrophysical background is predicted by the detailed analysis in~\cite{BringmannSalati}, the results of which we find to be well reproduced by a fitting function of the form 
${\rm log}_{10} \Phi_{\bar p}^{\rm bkg}$ = $-1.64 + 0.07\, \tau - \tau^2 - 0.02\, \tau^3 + 0.028\, \tau^4$
with $\tau = {\rm log}_{10} T/\GeV$. We take for definiteness the flux corresponding to the `MED' propagation parameters. Particularly favorable is the fact that the uncertainty in the estimates of the background is quite narrow around $10 - 100$ GeV, where results are expected soon.

\bigskip

Fig.\fig{antiprotons} shows the results for the final $\bar p$ flux from MDM annihilations at Earth (at the top of the atmosphere), normalized to the one of protons, compared to the background and to the most recent PAMELA data (older experimental data point are not anymore significant). We have assumed the same boost factor as for positrons, although in principle it could be different~\cite{Lavalle2}. As apparent, the agreement with the data is quite good. A prominent excess is predicted to show up at energies slightly above those that have been probed by PAMELA so far. This is a consequence of the large mass of the MDM candidate and also of the annihilation channel into $W^+W^-$.
The figure is produced with the benchmark choices of NFW and MED. The shape of the spectrum is relatively independent from the propagation model and the halo profile. Different $\bar p$ propagation models instead change the overall signal rate by about one orders of magnitude.
Different halo profiles with fixed $\rho_\odot$ make only a difference of a factor of a few, which can be interpreted in terms of the fact that the signal is not dominated by the far galactic center region, where profiles differ the most.

\subsection{Gamma rays}
Gamma ray signals from DM annihilation in the regions where DM is more dense can be very significant, but also depend strongly on the astrophysical assumptions, such as in particular the galactic DM profile. The spectra of high energy $\gamma$ rays from the galactic center in the Minimal DM case have been presented in~\cite{MDM2, MDM3}: the spectrum is characterized by a (smeared) line at $E \approx M$ and a continuum that extends to lower energies. Most astrophysical detections of gamma ray fluxes are however compatible with power law fluxes: the most conservative approach is therefore to assume that the observed fluxes are of standard (yet unknown) astrophysical origin and impose that the   flux from DM does not excede such observations. This imposes stringent constraints on the DM annihilation cross section and the DM halo profiles.

A complete, model independent calculation has been carried out in this respect in~\cite{BCST}, which considers bounds from high energy gamma rays from the galactic center, the galactic ridge and the Sagittarius dwarf galaxy, but also radio waves from the galactic center (produced by synchrotron radiation in the strong magnetic field by the electrons and positrons from DM annihilations). Such bounds apply of course in particular to the case of a particle with a 9.6 TeV mass and $W^+W^-$ main annihilation channel (fig.4 in~\cite{BCST}) such as Minimal DM: it is found that the annihilation cross section in eq.\eq{annihilationsigma} is (marginally) allowed in the case of a NFW DM profile in the Milky Way and the Sagittarius dwarf galaxy. It has however to be assumed that no boost factor is present for signals from the galactic center, which is possible if DM clumps are tidally disrupted in the central regions of the Milky Way. Alternatively, perhaps more realistically, but in tension with state-of-the-art numerical simulations, choosing less steep profiles such as isothermal allows to pass the constraints. 
Recent studies~\cite{Pieri,Strigari} along these lines for dwarf galaxies find comparable or looser constraints.

\section{Collider searches}
\label{collider}
At an accelerator like the Large Hadron Collider (LHC), which will soon(er or later) operate colliding $pp$ at $\sqrt{s} = 14 \TeV$, DM particles can in principle be pair produced, and this is a very promising way to search for DM and measure its properties. 

\medskip

In the case of MDM, for any multiplet with $Y=0$ and arbitrary $n$ the partonic total cross sections
(averaged over initial colors and spins) for producing any of its component are
\beq 
\hat\sigma_{u\bar d}= \hat\sigma_{d\bar u}=2\hat\sigma_{u\bar u} = 2\hat\sigma_{d\bar d}=
\frac{g_{\cal X} g_2^4(n^2-1)}{13824\ \pi \hat s}\beta \cdot\left\{\begin{array}{ll}
\beta^2 & \hbox{if ${\cal X}$ is a scalar}\\
3-\beta^2&\hbox{if ${\cal X}$ is a fermion}\end{array}\right.
\eeq
where  the subscripts denote the colliding partons,
and $\beta = \sqrt{1-4M^2/\hat{s}}$ is the  DM velocity with respect to
the partonic center of mass frame.
Production of non-relativistic scalars is $p$-wave suppressed in the usual way.
It is obvious that the production of the 5-plet with $M=9.6$ TeV is kinematically impossible.
Possible upgrades of LHC luminosity and magnets are discussed in~\cite{LHC28}.

If produced in collisions, MDM has a clean signature: the small mass splitting $\Delta M$ among the DM components makes charged MDM component(s) enough long-lived that they manifest  in the detector as charged tracks.
Irrespectively of the DM spin the life-time of $\DM^\pm$ particles with $Y=0$
and $n=\{3,5,7,\ldots\}$ is
$\tau \simeq 44\,{\rm cm}/(n^2-1)$
and the decay channels are precisely determined as 
\beq
\label{eq:gamma}
\begin{array}{lcll}
\DM^\pm \to \DM^0 \pi^\pm \qquad&:&\displaystyle
\Gamma_\pi =(n^2-1) \frac{G_{\rm F}^2V_{ud}^2\, \Delta M^3 f_\pi^2}{4\pi}
\sqrt{1-\frac{m_\pi^2}{\Delta M^2}},\qquad
& {\rm BR}_\pi =97.7\%\\[3mm]
\DM^\pm \to \DM^0 e^\pm\nubarnu_e &:& \displaystyle
\Gamma_e = (n^2-1) \frac{G_{\rm F}^2 \,\Delta M^5}{60\pi^3} &
{\rm BR}_e=2.05\%\\[3mm]
\DM^\pm \to \DM^0 \mu^\pm\nubarnu_\mu &:& \displaystyle
\Gamma_\mu =0.12\ \Gamma_e &
{\rm BR}_\mu=0.25\%\\
\end{array}
  \eeq
 having used the normalization $f_\pi = 131\,{\rm MeV}$~\cite{CHPT} and the $\Delta M$ of eq.\eq{166},
 which accidentally happens to be the value that maximizes ${\rm BR}_\pi$.
The $\DM^+$ life-time is long enough that decays can happen inside the detector.
  On the contrary, the faster decays of $\DM^{\pm\pm}$ particles (present for $n\ge 5$) mostly happen within the non-instrumented region with few cm size around the collision region.
Measurements of $\tau$ and of the energy of secondary soft pions, electrons and muons
constitute tests of the model, as these observables negligibly depend on the DM mass $M$.
On the contrary, measurements of the total number of events or of the
DM velocity distribution would allow to infer its mass $M$ and its spin.
Although SM backgrounds do not fully mimic the well defined MDM signal,
at an hadron collider (such as the LHC) their rate is so high that
it seems impossible to trigger on the MDM signal.

Notice that extra $\SU(2)_L$ multiplets that couple (almost) only through gauge interactions
tend to give a LHC phenomenology similar to the one discussed above,
irrespectively of their possible relevance for the MDM problem.
On the contrary, DM candidates like neutralinos are often dominantly produced
through gluino decays, such that
DM is accompanied by energetic jets rather than by charged tracks.

\section{Other phenomenological signatures}
\label{other}

The characteristic small mass splitting $\Delta M$ between the neutral DM and the charged components of the multiplet allows for a few other interesting albeit quite speculative manifestations of MDM.

\subsection{Accumulator}
\label{DMtron}

One could envision accelerating a large amount of protons $p$ or nuclei ${\cal N}$ in an accumulator ring and having them collide on the DM particles of the galactic halo bath, that would therefore act as a diffuse target. At energies $E_{p,{\cal N}} > \Delta M$ the CC collision $\DM \, {\cal N} \to {\cal N}^\pm \DM^\mp$ becomes kinematically possible and the DM$^\mp$ product would scatter out of the beam pipe giving a signature (collider experiments need instead a much larger energy $E_{p}\circa{>} M$ in order to pair produce DM). An estimate of the event rate gives
\beq 
\frac{dN}{dt}=\varepsilon N_p \sigma \frac{\rho_{\rm DM}}{M}=\varepsilon
\frac{10}{{\rm year}} \frac{N_p}{10^{20}} \frac{\rho_{\rm DM}}{0.3\GeV/{\rm cm}^3}\frac{\TeV}{M}
\frac{\sigma}{3\sigma_0}
\eeq
where $\rho_{\rm DM}$ is the local DM density and
$\varepsilon$ is the detection efficiency,
related e.g.\ to the fraction of beam that can be monitored. $\sigma_0 = G_{\rm F}^2 M_W^2/\pi=1.1\,10^{-34}\,{\rm cm}^2$ is a reference partonic cross section: at large energies $M_W\circa{<} E_p\circa{<} M$ the MDM-quark cross section approaches $3\sigma_0 (n^2-1)/4$. 
Proton drivers currently planned for neutrino beam experiments can produce
more than $10^{16}$ protons per second,
and accumulating $N_p\sim 10^{20}$ protons is considered as possible, so that the number of expected events looks not unreasonable.
The main problem however seems disentangling the signature from the beam-related backgrounds, such as collisions on residual gas.

\subsection{Ultra High Energy Cosmic Rays}
\label{UHECR}

\begin{figure}[t]
$$\hspace{-5mm}
\includegraphics[width=0.5\textwidth]{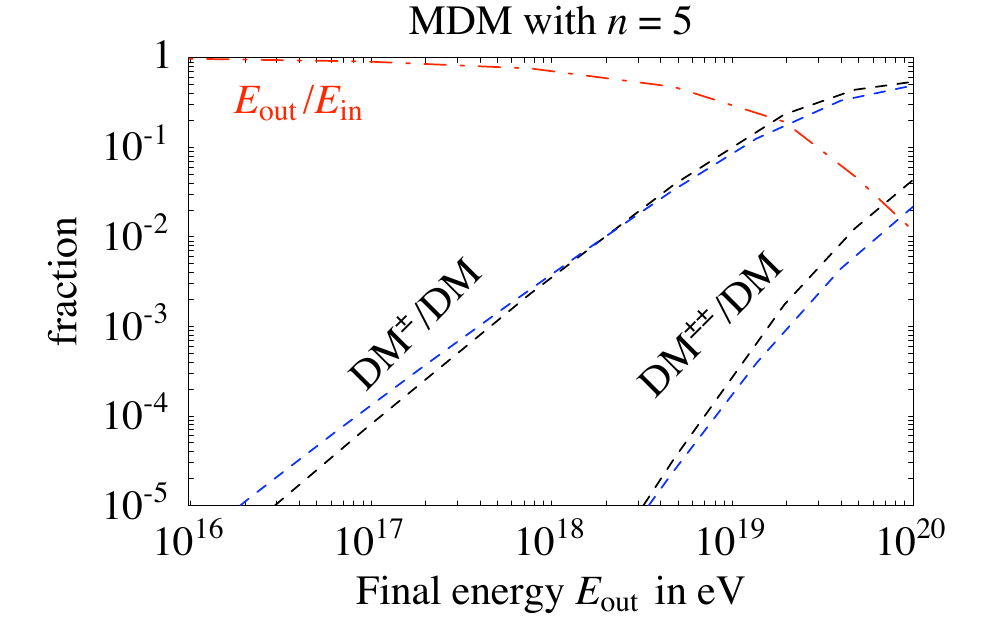}$$
\caption{\label{fig:UHECR}\em
{\bf MDM in UHECR}.  The fraction of DM particles that exit from an Earth-crossing journey in the charged states DM$^\pm$ and DM$^{\pm\pm}$ (dashed lines; the results of numerical and semi-analytical computations are represented by the paired lines that almost coincide), as a function of the exit energy $E_{\rm out}$. The red dot-dashed line show the ratio $E_{\rm out}/E_{\rm in}$ and allows to reconstruct the initial energy with which they must have entered the other side of the Earth. For instance, one reads that about 10\% of the DM particles that exit the Earth at $10^{19}$ eV do so in the DM$^\pm$ state; they have lost about 2/3 of their $E_{\rm in}$ in the journey. This fraction has to be multiplied by the flux of UHECR and by the fraction of it which is assumed to be made of MDM particles in order to get an actual event rate. UHECR have been detected up to few $\cdot 10^{20}$ eV.}
\end{figure}

If MDM particles are somehow a component of the Ultra High Energy (UHE) Cosmic Rays, the already relatively long half-life of DM$^\pm$ is Lorentz-dilated to macroscopical distances, and the possibility opens of seeing spectacular signatures in the form of km-long charged tracks of DM$^\pm$ in detectors like Icecube or Antares. 
Consider a flux of $\DM^0$ particles that is crossing the Earth at UHE. Via Charged Current (CC) interactions with nucleons of Earth's matter, $\DM^\pm$ particles are produced. Being charged particles traveling in a medium, these loose a part of their energy and eventually decay back to $\DM^0$ particles. 
This chain of production and decay is analogous to the process that tau neutrinos undergo in matter (``$\nu_\tau$-regeneration"). In the $n=5$ case $\DM^{\pm\pm}$ states are also produced and decay.
Such a system is described by a set of coupled integro-differential equations for the evolution of the fluxes of $\DM^0$, of $\DM^\pm$ and of $\DM^{\pm\pm}$ as a function of their energy and the position through the Earth.
At high energies, the CC cross section interactions are large (for $E\circa{>}10^{15}$ eV DM particles interact  on average at least one time along the whole diameter of the Earth), but so are also the energy losses and the Lorentz-dilated mean life of DM$^\pm$ and DM$^{\pm\pm}$, so that a detailed analysis (numerical or semi-analytical) is necessary.
The results are reported in fig.\fig{UHECR}b: the relevant phenomenological variable is the fraction of particles, at a given energy, that on average would emerge from their Earth crossing in the charged state and therefore would leave detectable tracks in the neutrino telescopes. One sees that, at high energies, such a fraction can be sizable. 
However, at those very high energies the overall flux of Cosmic Rays is not vanishing but fairly faint, so that the overall number of events would be limited.
Moreover, it seems challenging to construct a mechanism for accelerating MDM particles at UHECR energies in the first place. They could however be a component of cosmic rays generated by the decay of ultra-heavy particles, in the so called `top-down' scenario for the explanation of the origin of UHECR.  
Finally, particle identification in a detector like IceCUBE or Antares poses difficulties: $\DM^\pm$ would roughly look like muons (produced by neutrino interactions) with fake energy $E_\mu = E_\pm \beta_{\rm CC}/\beta_\mu \sim 10^{-4} E_\pm$,
because muon energy losses are approximatively given by
$-dE_\mu/d\ell = \alpha+\beta_\mu E_\mu$ with
$\beta_\mu \approx 0.2/{\rm kmwe}$~\cite{PDG} ($\beta$ is roughly inversely proportional to the particle mass).
One therefore needs to carefully study the energy loss profile along the charged tracks to
see the difference between a  $\DM^\pm$ with $E_\pm\sim 10^{18}\eV$
and a muon with $E_\mu\sim 10^{14}\eV$. Larger and more densely instrumented detectors seem necessary to study these issues.

\section{Conclusions and outlook}
\label{conclusions}

As experimental searches for Dark Matter proceed more vigorously than ever, many models may finally have the chance of being tested. Supersymmetric DM, Kaluza-Klein DM, Little Higgs DM etc.\ all make predictions that however 
(i) inscribe themselves into the context of oddly not-yet-found EW-scale New Physics;
(ii) typically depend on a large number of unknown model parameters, so that DM phenomenology remains dark; 
(iii) usually rely on the existence of some extra feature (R-, KK-, T-parity) imposed by hand in order to be kept stable against decay into SM particles.

The Minimal Dark Matter proposal takes on a more... minimal approach: focussing on the DM problem only, we add to the Standard Model just one extra multiplet ${\cal X}$ with electroweak interactions (such that it does not ruin the successful predictions that the SM makes in its renormalizable limit: conservation of baryon number, lepton number, lepton flavor, etc.), and investigate whether it can constitute a good DM candidate: electrically neutral, stable, produced in the right amount via the thermal freeze-out in the Early Universe and not excluded by direct DM detection results. 
We find that indeed the construction selects one preferred candidate, the fermionic $SU(2)_L$ quintuplet with hypercharge $Y=0$, the main phenomenological properties of which are listed in Table \ref{tab:2}. 
These properties are univocally computed, as no free parameters are present in the model. 
Also, no extra feature is introduced by hand to guarantee its stability: simply no decay modes exist consistently with the SM gauge symmetry (analogously to the stability of the proton in the SM).
Phenomenological predictions can be univocally listed:
\begin{table}[t]
\center
\begin{tabular}{|c|c|c|c|cc|}
\hline
`definition' & DM mass   & splitting & Direct Detection & \multicolumn{2}{c|}{\hbox{Galactic annihilation}}  \\
 & $M$ & $\Delta M$ & $\sigma_{\rm SI}$ & $\langle \sigma v \rangle$ & channel \\[1mm]
\hline
$\begin{array}{c}
{\rm fermionic} \\
$SU(2)$_L \ 5\hbox{\rm-plet}
\end{array}$ & $9.6$ TeV & $166$ MeV & $1.2 \cdot 10^{-44}$ cm$^2$ & $1.1 \cdot 10^{-23} \ \frac{{\rm cm}^3}{{\rm sec}}$  & $W^+W^-$ \\
$Y = 0$ & &  &  & $3.3 \cdot 10^{-24}  \ \frac{{\rm cm}^3}{{\rm sec}}$ & $Z Z $ \\[1mm]
 & & & & $2.0 \cdot 10^{-24}  \ \frac{{\rm cm}^3}{{\rm sec}}$ & $\gamma Z $ \\[1mm]
  & & & & $3.0 \cdot 10^{-25}  \ \frac{{\rm cm}^3}{{\rm sec}}$  & $\gamma\gamma$ \\[1mm]
\hline
\end{tabular}
\caption{\label{tab:2} \em Main phenomenological properties of the preferred MDM candidate. Estimates of the theory uncertainties on the numerical values are given in the text.}
\end{table}

\begin{itemize}

\item[$-$] The univocal MDM signature at {\bf colliders} is production of $\DM^\pm$, 
that manifests as a non-relativistic charged track (straight despite the magnetic field $B\sim {\rm Tesla}$ in the detector)
that decays with a relatively long life-time of $\tau \simeq 1.8$ cm (due to the small splitting) into $\DM^0\pi^\pm$, leaving a quasi-relativistic curved track.
However, this signature will not be visible at LHC, because the mass of the preferred MDM candidate is too large
(see~\cite{LHC28} for possible LHC upgrades) and because this signature is too complex for
triggers (given their speed and the rate of QCD backgrounds).

\item[+] The next generation of {\bf direct detection} experiments, such as Super-CDMS or Xenon 1-ton, prospect to be sensitive to the spin-independent scattering cross section of the MDM candidate on nuclei, see fig.\fig{direct}. 
MDM  cannot account for the controversial annual modulation claimed by DAMA/Libra.

	\item[+]  {\bf Indirect DM searches} constitute arguably the most interesting testing ground for the MDM model. The predictions of the theory had been presented in~\cite{MDM3} and can now be confronted to the recent experimental results from PAMELA, ATIC and HESS, of course under the assumption that these are interpreted in terms of DM annihilations. The main features of the predicted spectra are determined by the following main properties of the model:
\begin{itemize}
\item[$\cdot$] the DM mass is very large (9.6 TeV), so that fluxes are expected to extend to multi-TeV energies;
\item[$\cdot$] the predominant annihilation channel is into $W^+W^-$ channel, that produces fluxes of all species ($e^+$, $\bar p$, $\gamma$ and $\nu$) with a characteristic spectrum;
\item[$\cdot$] the total annihilation cross section is very large (of the order of $10^{-23} \ {\rm cm}^3/{\rm sec}$) thanks to the non-perturbative Sommereld enhancement which is present due to the exchange of EW gauge bosons between the annihilating particles, as reviewed in Appendix A.
\end{itemize}
The profile of DM distribution in the galaxy and the propagation of the fluxes of charged anti-matter in the galaxy introduce an uncertainty on the final spectra that is however of limited impact, essentially because the large mass of the MDM particle produces fluxes of high energy final products that do not travel for long in the galaxy otherwise they would loose all their energy (the dependence on the propagation parameters remains more significant, at the level of one order of magnitude, on $\bar p$ fluxes). 
Fig.s\fig{positrons}, \fig{leptons} and \fig{antiprotons} illustrate that: 
\begin{itemize}
\item[+] The MDM 5-plet prediction agrees with the {\bf PAMELA data on the positron fraction}, if an astrophysical boost factor of $B \simeq 50$ is introduced (reduced to $B \simeq 5$ within the $3\sigma$ allowed uncertainty range for the annihilation cross section). The model predicts a continuous rise up to an energy corresponding to the DM mass of about 10 TeV.   
\item[+] The MDM 5-plet prediction agrees with the {\bf PAMELA data on antiproton/proton fluxes}, assuming the same astrophysical boost factor. The reason of the agreement is that the excess in $\bar p$ corresponding to the excess in $e^+$ starts at energies slightly higher than those probed by PAMELA so far. Therefore the model predicts a very relevant anomaly in future PAMELA or AMS-02 data with respect to the expected background. 
\item[$-$] The MDM 5-plet flux of the sum of electrons and positrons is {\bf not} compatible with the {\bf peak suggested by ATIC} and PPB-BETS data at around 600 GeV: an $e^++e^-$ spectrum which is flat up to larger energies is expected instead. 
The HESS observations hint to a steepening in the $e^++e^-$ spectrum, without however testing the peak.
{\em New data should soon clarify the experimental situation, possibly falsifying MDM.}
\footnote{{\bf Note added}.
The ATIC peak, incompatible with MDM predictions, is now contradicted by the new FERMI data~\cite{FERMI} (unless an exceptionally bad energy resolution is assumed for FERMI).
However FERMI data are consistent with the steepening apparent in the HESS data (recently supplemented by the new results at lower energy in~\cite{HESS2009}).}
\end{itemize}
Moreover:
\begin{itemize}
\item[$\sim$] The production of gamma rays (and of radio synchrotron emission from the $e^\pm$) from the MDM 5-plet at the galactic center and in dwarf galaxies are compatible with the observations for an NFW or Einasto profile if the astrophysical boost is assumed not to be present for observations in the gamma channel towards the galactic center, or if a not-too-steep DM profile such as isothermal is assumed (somewhat in tension with numerical simulations).
\item[$\sim$]Neutrinos from the annihilations of MDM trapped at the Sun's center are not expected to be detectable, as the fluxes depend on the inverse of the square of the DM mass (which is large) and the spin dependent capture cross section on solar nuclei is small. Neutrinos from the galactic center can be relevant for detectors with sensitivity larger than the current one by a few orders of magnitude. 
\item[$\sim$] Anti-deuterium $\bar d$ is formed when the DM annihilation products contain a $\bar p$ and a $\bar n$ with momentum difference (in their rest frame) less than about 160 MeV~\cite{Dbar}. Ref.~\cite{DMdeuterium1} computed the $\bar d$ spectrum from MDM in the standard spherical approximation~\cite{Dbar}, finding a yield well below the background. However Minimal Dark Matter annihilates into $W$ boosted by a Lorentz factor $\gamma = M/M_W\approx 120$, and the jet structure of the products has to be taken into account. This has been done in~\cite{DMdeuterium2}. The total $\bar d$ yield is the same as the one from decays of a $W$ at rest, but everything is boosted by a factor $\gamma$, resulting (for the central value of $\sigma v$ we assume) into a $\bar d$ flux of about $10^{-10}$ per m$^2,{\rm sec}\,{\rm sr}\,{\rm GeV}$ at energies of about 100 GeV per nucleon. This should be then comparable to or above the expected astrophysical $\bar d$ background. These large energies are however challenging for the purposes of detection.
\end{itemize}
If the anomalies in the fluxes of leptons ($e^+$ fraction and $e^++e^-$) in the ranges of energies currently explored will turn out to be of astrophysical origin, e.g. from one or more local pulsars~\cite{pulsars,astro}, the predictions of MDM (as well as of most DM models) will be drowned in an unsuppressible background. The MDM model is peculiar in predicting, however, excesses that continue up to 10 TeV in these channels.  Also, the prediction for the flux of anti-protons (much more difficult to mimic with astrophysics) would stay, and it cannot be reduced below a minimum level: with unit boost factor, the MDM flux is still marginally above the background for most choices of parameters, albeit concentrated at very high energies.
\item[+] Constraints on the amount of energy deposited by DM annihilations at the time of BBN (when the velocity of the DM particles can be $\circa{<}10^{-3}$) impose an upper bound on the annihilation cross section. They were recently reconsidered by~\cite{HisanoBBN} after~\cite{Jedamzik}: the annihilation cross section of MDM eq.\eq{annihilationsigma} is found to be allowed (see e.g. fig. 2 of~\cite{HisanoBBN}; note that for MDM $\sigma v$ is already at saturation at $\beta \simeq 10^{-3}$, see fig.\fig{relic}b).
\end{itemize}
Further experimental results are expected soon.

\appendix

\section{Sommerfeld effects}
The DM DM annihilation cross section gets enhanced if DM particles have a non-relativistic velocity $\beta \ll 1$
and there is a long-range attractive force between them.
Technically, the effect arises because annihilating DM particles  cannot be approximated as plane waves,
so their wave-function must be computed by solving the Schroedinger equation in presence of the long-range potential $V$.
For a single abelian massless vector with potential $V=\alpha/r$ the result is
$\sigma = S \sigma_{\rm perturbative}$ where the Sommerfeld correction is~\cite{Sommerfeld}
\beq  \label{eq:R0}
{\cal S}(x) = \frac{-\pi x}{1-e^{\pi x}} \qquad x = \frac{\alpha}{\beta}.\eeq
Here $\alpha<0$ describes an attractive potential that leads to an enhancement ${\cal S}> 1$, and
$\alpha>0$ describes a repulsive potential that leads to ${\cal S}<1$.
The above results holds for the $s$-wave $(L=0)$ partial wave, that dominates in the non-relativistic limit:
the $s$-wave annihilation cross section $\sigma \beta$ roughly grows as $1/\beta$ for
$\beta\circa{<}\alpha$.
Higher waves ($L>0$) are enhanced by higher powers of $\alpha/\beta$~\cite{Iengo} such that, as $\beta \to 0$ all partial cross sections
grow as $1/\beta$, but with negligible coefficients suppressed by $\alpha^{2L}$.

\smallskip

The Sommerfeld enhancement is automatically present in the MDM case, since the MDM particle is coupled to the SM $W,Z$ vectors and it is significantly heavier than them.
As a consequence the $s$-wave annihilation cross section $\sigma \beta$ grows as $1/\beta$ 
for $1\circa{>} \beta \circa{>} M_{W}/M \sim 10^{-3}$. This is why a possibly large positron excess had been predicted in the MDM framework before the PAMELA results.\footnote{In other DM models, ad hoc light new particles are often introduced such that the Sommerfeld enhancement allows to accomodate the PAMELA anomaly (as it suggests a $\sigma \beta$ at $\beta \sim 10^{-3}$ which is a few orders of magnitude larger than 
the $\sigma \beta \approx 3~10^{26}{\rm cm}^3/{\rm sec}$ at $\beta \sim 0.2$
suggested by the cosmological DM abundance).}

In the MDM framework the force vectors (the weak gauge bosons) are massive and non-abelian:
 the computation of the Sommerfeld effect is more involved as one must classify the two-body DM DM states according to their conserved quantum numbers: angular momentum $S$,
total spin $L$ and total electric charge $Q$. Ref.~\cite{MDM2} presents all the details of the calculations in the different cases.
Here we focus on the $\DM^0\DM^0$ state of the 5-plet, relevant for astrophysical signals, which lies in the $Q=0$, $L=0$, $S=0$ sector containing 3 states: $\{\DM^{++}\DM^{--}, \DM^+\DM^-,\DM^0\DM^0\}$.
The `potential' $V$ and the `annihilation rate' $\Gamma$ become $3\times 3$ matrices,
whose off-diagonal components  (and especially their signs) do not have an intuitive meaning and are defined
in terms of the real and imaginary part of the two-body matrix propagators.
The explicit result is~\cite{MDM2}
\beq
V = \bordermatrix{&--&-&0\cr
++ &8\Delta -4A &- 2B &0\cr
+&-2B& 2\Delta - A&
-3\sqrt{2} B\cr
0&0&-3\sqrt{2} B & 0},\qquad
 \Gamma=\frac{3\pi\alpha_2^2}{25M^2}\bordermatrix{&++&+&0\cr
-- & 12 &6 &2\sqrt{2}\cr
- &6 &9&5\sqrt{2}\cr
0 &2\sqrt{2}&5\sqrt{2}&6},
\eeq
where $\Delta=166\MeV$,
$A =   \alpha_{\rm em}/r + \alpha_2 c_{\rm W}^2 e^{-M_Zr}/r$ and 
$B=\alpha_2e^{-M_Wr}/r$.
The Sommerfeld enhancement can now be computed by numerically solving the  matricial Schroedinger equation.

\medskip

Co-annihilation with all other DM components enter in the computation of the cosmological freeze-out abundance,
so that a lengthy computation is needed. We here discuss how one
can perform a simplified computation in the limit of unbroken SU(2)$_L$,  neglecting the SM vector masses $M_{W,Z}$
and the MDM intra-multiplet mass splitting $\Delta$.  This approximation is good enough for the cosmological computation.
Group theory allows to reduce the unbroken non-abelian Sommerfeld enhancement to a combination of abelian-like
enhancements in the various sectors~\cite{SomNonAb}, classified according to
the conserved quantum numbers: $L,S$ and iso-spin $I$.
We can focus on $s$-wave annihilations $(L=0)$ and on states with $I\le 5$, that  annihilate at tree level into two SM particles.
The Pauli principle restricts the allowed states to be $(I,S)=(3,1)$, $(1,0)$ and $(5,0)$.
After performing the relevant thermal average, the $s$-wave coefficient of eq.\eq{cs}
gets Sommerfeld-enhanced to
\beq \label{eq:SF} c_s =   
\underbrace{\frac{30 g_2^4}{\pi} {\cal S}(-5\alpha_2 \sqrt{M/T})}_{(I,S)=(1,0)\to W^a W^a} + 
\underbrace{\frac{105 g_2^4}{2\pi} {\cal S}(-3\alpha_2 \sqrt{M/T})}_{(I,S)=(5,0)\to W^a W^b} +
\underbrace{(1+\frac{1}{24})
\frac{45g_2^4}{\pi}{\cal S}(-6\alpha_2\sqrt{M/T})}_{(I,S)=(3,1)\to \Psi\Psi,HH}.
 \eeq
The resulting DM freeze-out abundance is shown in the left panel of fig.\fig{relic} (dashed line), compared with the full computation (continuous line).

\section{Next-to-minimal Minimal Dark Matter}
In this Appendix we briefly sketch the directions into which the model can be extended, if the request of full minimality is abandoned and some tools of DM model building (commonly used elsewhere) are introduced.

\medskip

In Sec.\ref{model} we rejected the candidates which have dimension-4 or dimension-5 interactions with SM fields and can therefore decay into SM particles in a time much shorter than the age of the universe. By invoking some ad hoc extra symmetry that prevents the decays, these candidates can of course be reallowed. 
This is actually also what happens in the case of the best know WIMP DM candidates: in SUSY models, by assigning odd $R$-parity to the supersymmetric partners including the LSP DM, the decay operators, that feature just one DM state, are forbidden. An equivalent mechanism is used in little-Higgs models ($T$-parity), in `universal' extra dimension models (KK-parity), etc.
Some of the MDM candidates which are rescued by the imposition of the extra symmetry resemble to particles that appear  in a variety of other contexts: e.g.\ scalar triplets in little-Higgs models; fermion or scalar triplet in see-saw models; KK excitations of lepton doublets or of higgses in extra dimensional models; higgsinos, sneutrinos, winos in supersymmetric models. 
In general, however, these models possess a plethora of other parameters so that typically their DM candidates behave very differently from the MDM candidates. But in some corner of the parameter space the phenomenological features can coincide. The signatures of some of these candidates (in particular the `wino-like' fermionic triplet and the scalar triplet) have been studied in~\cite{MDM2, MDM3}.

Speaking of decays, we note that the main MDM candidate (the fermion 5-plet, cosmologically stable in the minimal construction) acquires a non-negligible decay mode if $\Lambda$ is somewhat below the Planck scale.
The effective dimension-6 operator that dominantly contributes to the decay is ${\cal X} L HH H^*/\Lambda^2$. 
It induces 4-body decays, such as  $\DM^0 \to \ell^\mp W^\pm_L Z_LZ_L$
(where $L$ denotes longitudinal polarization), with rate $\Gamma \sim M^5/4\pi \Lambda^4$.
Such decays can provide an alternative interpretation of the PAMELA excess: $\Lambda\approx 10^{16}\GeV$ gives a life-time of about $10^{25}\,{\rm sec}$.
2-body decays, such as $\DM^0 \to \ell^\pm W^\mp$ have a rate $\Gamma_2 \sim v^4M/(4\pi)^5 \Lambda^4$
which is subdominant as long as $M\gg 4\pi v$.

\medskip

As discussed in Sec.\ref{direct}, MDM candidates with $Y\neq 0$ have an elastic cross section which is
 $2\div 3$ orders of magnitude above present direct detection bounds~\cite{CDMS} due to the exchange of a $Z$ boson and have therefore been rejected. A possible way to reinclude them is to abandon minimality by introducing another state that mixes with the DM: this has the effect of splitting the components of the Dirac DM particle by an amount $\delta m$ such that the lightest one becomes a Majorana fermion which cannot have a vector-like coupling to the $Z$ boson. This is actually what happens in the well-known case of SuSy Higgsino DM, via the mixing with Majorana gauginos. In our case, if a $\delta m$ larger than the DM kinetic energy and smaller than $M_Q-M_0$ is generated, the only consequence would be to suppress the direct detection DM signals and the rest of the phenomenology would be essentially unchanged.

\medskip

In the case of scalar DM, additional operators not listed in eq.\eq{lagrangian} are generically present: 4-scalar interactions of ${\cal X}$ with higgses and ${\cal X}$ quartic self-interactions:
\beq  
\Lag \supset -c\, \lambda_H ({\cal X}^*  T^a_{\cal X} {\cal X})\, (H^* T^a_H H)-
c\, \lambda'_H |{\cal X}|^2 |H|^2 -   \label{eq:Lnonminimal}
 \frac{\lambda_{\cal X}}{2} ({\cal X}^*  T^a_{\cal X} {\cal X})^2 -  \frac{\lambda'_{\cal X}}{2} |{\cal X}|^4,
 \label{eq:non-minimal}
 \eeq
where $T^a_{R}$ are $\SU(2)_L$ generators in the representation to which ${R}$ belongs and $c = 1~(1/2)$ for a complex (real) scalar. In the strictly minimal theory these terms have been assumed vanishing, as they introduce the new unknown couplings $\lambda_H$, $\lambda'_H$, $\lambda_{\cal X}$ and $\lambda'_{\cal X}$. But if minimality is relaxed they can be introduced and they allow to enlarge the parameter space of the theory.
These extra interactions do not induce DM decay (because two ${\cal X}$ are involved, and we assume $\langle {\cal X}\rangle = 0$).
They affect instead the computations of the mass splitting, of the direct detection inelastic cross section and of the relic density (therefore the determination of the DM mass).
Indeed, the coupling $\lambda_H$ splits the masses of the components of ${\cal X}$ by an amount
\beq 
\label{eq:tree} 
\Delta M =\frac{ \lambda_H v^2 |\Delta T^3_{\cal X}|}{4M}  = \lambda_H \cdot 7.6\GeV \frac{\TeV}{M}
\eeq
having inserted $\langle H\rangle  = (v,0)$ with $v=174\GeV$
and $\Delta T^3_{\cal X} = 1$.
This cannot be neglected with respect to the splitting which is separately generated by loop corrections (discussed in Sec.~\ref{splitting}) as soon as $\lambda_H \approx 0.01$.
Also, $\lambda_H$ and $\lambda'_H$ open the possibility of extra annihilations into higgses
\beq 
\label{eq:extra}
\langle \sigma_A v\rangle_{{\rm extra}} = 
\frac{|\lambda_H^{\prime}|^2 + (n^2-1) |\lambda_H|^2/16}{16\pi\ M^2\ g_{\cal X} }
\eeq
(no interference terms are present).
The contribution from $\lambda_H$ is relevant for $\lambda_H \approx 0.01$, the same value that also produce a non-negligible mass splitting. The contribution  from $\lambda'_H$ is relevant for $|\lambda'_H| \sim g_{Y}^2,g_2^2$ or larger. 
In these regimes, to compute how much the inferred value of $M$ would be affected one simply adds the contribution in eq.\eq{extra} to those in eq.\eq{sigAs}: in general, the increase of the total annihilation cross section implies the increase of the inferred MDM mass, and the mass of the low $n$ candidates are more affected than the high $n$ ones, because the gauge-mediated annihilation cross sections are relatively less important for the formers. 
For example, a large $\lambda'_H = 1$ would increase the predicted value of
$M$ by a factor $2.4$ for $n=2$, by $20\%$ for $n=3$, by $2\%$ for $n=5$,
and by $0.5\%$ for $n=7$.
Note that $\lambda_H \sim g_2^2$ and $\lambda'_H \sim g_Y^2$ are the values
predicted by some solutions to the hierarchy problem, such as supersymmetry and gauge/higgs unification.
Finally, non-zero couplings $\lambda_H$ and $\lambda'_H$ can also produce an elastic cross section $\sigma_{\rm SI}$ on nuclei, much larger than the one in eq.(\ref{direct}), as an UV-divergent effect that corresponds to a renormalization of $|{\cal X}|^2|H|^2$ operators is generated.

We note that  when ${\cal X}$ is a neutral scalar singlet,  the non-minimal annihilations in eq.\eq{Lnonminimal} are the only existing ones. In this case, the observed amount of DM is obtained for $M\approx 2.2\TeV|\lambda'_H|$ (we are assuming $M\gg M_Z$; for generic values of $M$ the correlation between  $M$ and $\lambda'_H$ was studied in~\cite{singlet}). 

\medskip

Finally, one can envision a scenario in which more than one MDM multiplet is present at the same time. Or also a scenario in which several distinct copies (`flavors', in analogy with the SM families) of the same MDM multiplet exist. 
In both cases, to a good approximation their abundances evolve independently, such that the observed DM abundance is reproduced for lower values of the respective DM masses. In this more general situation, the $M$ values in table~\ref{tab:1} must be reinterpreted as upper bounds on $M$. This might bring some of the candidates into the reach of the LHC.

\paragraph{Acknowledgements}
We thank our collaborators Nicolao Fornengo, Matteo Tamburini, Roberto Franceschini, Carolin Br\"auninger and Paolo Panci. We thank the EU Marie Curie Research \& Training network ``UniverseNet" (MRTN-CT-2006-035863) for support. We thank Daniel Feldman for pointing out a typo (here corrected) in Table 2.

\bigskip
\appendix

\footnotesize
\begin{multicols}{2}
  
\end{multicols}


\begin{thebibliography}{nn}


\bibitem{reviews}
Recent reviews include: 
G.~Bertone, D.~Hooper and J.~Silk,
  Phys.\ Rept.\  405 (2005) 279
  [arXiv:hep-ph/0404175].
J.~Einasto,
  arXiv:0901.0632 [astro-ph.CO].

\bibitem{FT} The fine-tuning argument was invented to  motivate new physics at the weak scale.
  Its back-firing was addressed, after LEP2 negative searches, in
  \art[hep-ph/9811386]{L. Giusti, A. Romanino, A. Strumia}{Nucl. Phys.}{B550}{3}{1999};
  \art[hep-ph/9905281]{R. Barbieri, A. Strumia}{Phys. Lett.}{B462}{144}{1999}.

\bibitem{MDM}
M.~Cirelli, N.~Fornengo and A.~Strumia,
  Nucl.\ Phys.\  B 753 (2006) 178
  [arXiv:hep-ph/0512090].

\bibitem{MDM2}
M.~Cirelli, A.~Strumia and M.~Tamburini,
  Nucl.\ Phys.\  B 787 (2007) 152
  [arXiv:0706.4071].

\bibitem{MDM3}
M.~Cirelli, R.~Franceschini and A.~Strumia,
  Nucl.\ Phys.\  B 800 (2008) 204
  [arXiv:0802.3378].

\bibitem{MDMidm08}
 M.~Cirelli and A.~Strumia,
  arXiv:0808.3867.

\bibitem{strong}
Strongly interacting, ``hadronized", Dark Matter is subject to a number of constraints analysed in \art{G.~D.~Starkman, A.~Gould, R.~Esmailzadeh and S.~Dimopoulos}{\PR}{D41}{3594}{1990} that leave only implausible windows open. Recently, \hepart[0705.4298]{G. D. Mack, J. F. Beacom and G. Bertone} went in the direction of closing those windows to safeguard Earth's heat flow.

\bibitem{GoodWit}
  M.~W.~Goodman and E.~Witten,
  Phys.\ Rev.\ D {31} (1985) 3059.

\bibitem{CDMS}
Z.~Ahmed {\it et al.}  [CDMS Collaboration],
  Phys.\ Rev.\ Lett.\  102 (2009) 011301
  [arXiv:0802.3530].
  
  \bibitem{Xenon}
  J.~Angle {\it et al.}  [XENON Collaboration],
  Phys.\ Rev.\ Lett.\  100 (2008) 021303
  [arXiv:0706.0039].


\bibitem{cosmoDM}
These numbers summarize 
various recent global analyses of cosmological data
within the $\Lambda$CDM model
that found compatible values and uncertainties:
\hepart[astro-ph/0603449]{D. N. Spergel {\it et al.} [WMAP collaboration]},
\art[astro-ph/0607086]{M. Cirelli and A. Strumia}{JCAP}{0612}{013}{2006}, 
\art[astro-ph/0608632]{M. Tegmark et al.}{Phys. Rev.}{D74}{123507}{2006}.

\bibitem{Coulomb}
C.A. Coulomb, ``Premier M\'emoire sur l'\'{E}le\-ctri\-cit\'e et le
Magn\'etisme'', M\'emoires de l'Acad\'emie Royale des Sciences, vol.88 (1785), 569.
  
\bibitem{DAMA}
 R.~Bernabei {\it et al.}  [DAMA Collaboration],
  Eur.\ Phys.\ J.\  C  56 (2008) 333
  [arXiv:0804.2741].
  
  \bibitem{Drees}
  \art{M. Drees, M. Nojiri}{\PR}{D48}{3483}{1993}.

\bibitem{Pittel}
  See also \art{J. Engel, S. Pittel, P. Vogel}{Int. Journ. Mod. Phys.}{E1}{1}{1992}.

\bibitem{future}
For the XENON project see:
E.~Aprile, L.~Baudis for the XENON Collaboration,
  arXiv:0902.4253 [astro-ph.IM],
proceeding of the {\it Identification of Dark Matter 2008} conference, published by {\it Proceedings of Science}.

For the SuperCDMS project see: \art[astro-ph/0503583]{CDMS-II collaboration}{
eConf}{C041213}{2529}{2004}.

Useful comparisons can be done using the tools in \url{dendera.berkeley.edu/plotter/entryform.html}.

\bibitem{PAMELApositrons}
O.~Adriani {\it et al.}  [PAMELA Collaboration],
  arXiv:0810.4995.
  
\bibitem{HEAT}
S.~W.~Barwick {\it et al.}  [HEAT Collaboration],
  Astrophys.\ J.\  482 (1997) L191
  [arXiv:astro-ph/9703192].

\bibitem{AMS01}
AMS-01 Collaboration: \art[astro-ph/0703154]{M. Aguilar et al.}{Phys. Lett.}{B646}{145-154}{2007}.

 \bibitem{PAMELApbar}
 O.~Adriani {\it et al.},
  arXiv:0810.4994.
  
  \bibitem{ATIC-2}
ATIC collaboration, Nature 456 (2008) 362.

\bibitem{Torii:2008xu}
\hepart[0809.0760]{PPB-BETS collaboration}. Web page: \url{http://ppb.nipr.ac.jp}.

\bibitem{HESSleptons}
 F.~Aharonian {\it et al.}  [H.E.S.S. Collaboration],
  Phys.\ Rev.\ Lett.\  101 (2008) 261104
  [arXiv:0811.3894].

\bibitem{ATIC-4} 
Talk by J.P. Wefel at ISCRA 2008, Erice, Italy, 2008,
\url{http://laspace.lsu.edu/ISCRA/ISCRA2008}.

\bibitem{EC}
T. Kobayashi, J. Nishimura, Y. Komori, T. Shirai, N. Tateyama, T. Taira, K. Yoshida, T. Yuda (EC collaboration), Proceedings of 1999 ICRC, Salt Lake City 1999, Cosmic ray, vol. 3, pag.\ 61--64.

\bibitem{FERMIleptons}
{\sc FERMI} collaboration, `{\em Measuring 10-1000 GeV Cosmic Ray Electrons with GLAST/LAT}',
talk at the ICRC07 conference.

\bibitem{CKRS}
M.~Cirelli, M.~Kadastik, M.~Raidal and A.~Strumia,
  arXiv:0809.2409.

\bibitem{DiffusionCylinder}
The now standard ``two-zone diffusion model" introduced in 
\art{V.L. Ginzburg, Ya.M. Khazan, V.S. Ptuskin}{Astrophysics and Space Science}{68}{295-314}{1980},  
\art{W.R. Webber, M.A. Lee, M. Gupta}{Astrophysical Journal}{390}{96-104}{1992}.

\bibitem{FornengoDec2007}
T.~Delahaye, R.~Lineros, F.~Donato, N.~Fornengo and P.~Salati,
  Phys.\ Rev.\  D 77 (2008) 063527
  [arXiv:0712.2312].
  
\bibitem{HisanoAntiparticles}
\art[hep-ph/0511118]{J. Hisano, S. Matsumoto, O. Saito, M. Senami}{Phys. \
Rev.}{D73}{055004}{2006}.

\bibitem{isoT}
\art{J. N. Bahcall and R. M. Soneira}{Astrophys. J. Suppl.}{44}{73}{1980}

\bibitem{Einasto}
 A.~W.~Graham, D.~Merritt, B.~Moore, J.~Diemand and B.~Terzic,
  Astron.\ J.\   132 (2006) 2685
  [arXiv:astro-ph/0509417].
\hepart[0810.1522]{J.~F.~Navarro {\it et al.}}.

\bibitem{NFW}
\art[astro-ph/9611107]{J. Navarro, C. Frenk, S. White}{Astrophys. J.}{490}{493}{1997}.

\bibitem{Moore}
\art[astro-ph/0402267]{J.~Diemand, B.~Moore and J.~Stadel}{Mon.\ Not.\ Roy.\ Astron.\ Soc.}{353}{624}{2004}.

\bibitem{DonatoPRD69}
 \art[astro-ph/0306207]{F.~Donato, N.~Fornengo, D.~Maurin and P.~Salati}{Phys.\ Rev.}{D 69}{ 063501}{2004}.

\bibitem{MoskalenkoStrong}
\art[astro-ph/9710124]{I.~V.~Moskalenko and A.~W.~Strong}{Astrophys.\ J.}{493}{694}{1998}.

\bibitem{bkgpositrons}
\art[astro-ph/9808243]{E.~A.~Baltz and J.~Edsjo}{Phys.\ Rev.}{D59}{023511}{1999}.

\bibitem{Delahaye}
T.~Delahaye, F.~Donato, N.~Fornengo, J.~Lavalle, R.~Lineros, P.~Salati and R.~Taillet,
  arXiv:0809.5268.
  

\bibitem{Lavalle1}
\hepart[astro-ph/0603796]{J.~Lavalle, J.~Pochon, P.~Salati and R.~Taillet}

\bibitem{Lavalle2}
\hepart[0709.3634]{J.~Lavalle, Q.~Yuan, D.~Maurin and X.~J.~Bi}.

\bibitem{minispikes}
\art[0704.2543]{P.~Brun, G.~Bertone, J.~Lavalle, P.~Salati and R.~Taillet}{Phys.\ Rev.}{D76}{083506}{2007}. 
  See however the recent 
  T.~Bringmann, J.~Lavalle and P.~Salati,
  arXiv:0902.3665.

\bibitem{Berez}
\art[astro-ph/0301551]{V.~Berezinsky, V.~Dokuchaev and Y.~Eroshenko}{Phys.\ Rev.}{D68}{103003}{2003}. 

\bibitem{crosssection}
\art{L.~C.~Tan and L.~K.~Ng}{J.\ Phys.}{G 9}{227}{1983}.

\bibitem{methodPbar}
\art[astro-ph/9606174]{P.~Chardonnet, G.~Mignola, P.~Salati and R.~Taillet}{Phys.\ Lett.}{B 384}{161}{1996}.
\art[astro-ph/9804137]{A.~Bottino, F.~Donato, N.~Fornengo and P.~Salati}{Phys.\ Rev.}{D 58}{123503}{1998}.
  See also: 
\art[astro-ph/9902012]{L.~Bergstrom, J.~Edsjo and P.~Ullio}{Astrophys.\ J.}{526}{215}{1999}.

\bibitem{TailletRRDA}  
\hepart[astro-ph/0212111]{D.~Maurin, R.~Taillet, F.~Donato, P.~Salati, A.~Barrau and G.~Boudoul}.
  
    \bibitem{MaurinApJ555}
\art[astro-ph/0101231]{D.~Maurin, F.~Donato, R.~Taillet and P.~Salati}{Astrophys.\ J.}{555}{585}{2001}.

\bibitem{GA}
\art{L.J. Gleeson and W.I. Axford}{ApJ}{149}{L115}{1967} and 
\art{L.J. Gleeson and W.I. Axford}{ApJ}{154}{1011}{1968}.
  
\bibitem{BringmannSalati}
\art[astro-ph/0612514]{T.~Bringmann and P.~Salati}{Phys.\ Rev.}{D 75}{083006}{2007}.
  
 \bibitem{BCST}
   G.~Bertone, M.~Cirelli, A.~Strumia and M.~Taoso,
  arXiv:0811.3744.
      
  \bibitem{Pieri}
  L.Pieri, M.Lattanzi and J.Silk, 
  arXiv:0902.4330.
 
   \bibitem{Strigari}
  R.~Essig et al.,
  arXiv:0902.4750.

 \bibitem{LHC28}
For discussions about realistic LHC upgrades see
\art{G. Azuelos et al. ({\sc Atlas} collaboration)}{J. Phys.}{G28}{2453}{2002};
A. de Roeck,  talk at the
2003 International Workshop on Future Hadron Colliders,
web site \url{conferences.fnal.gov/hadroncollider}.
D. Denegri, talk at the 2005 Les Houches Workshop,
web site \url{lappweb.in2p3.fr/conferences/LesHouches}.

\bibitem{CHPT}
For a recent review of chiral perturbation theory see
  G.~Colangelo and G.~Isidori, hep-ph/0101264.
 
 \bibitem{Dbar}
F.~Donato, N.~Fornengo and P.~Salati,
  Phys.\ Rev.\  D 62 (2000) 043003
  [arXiv:hep-ph/9904481].
 
 \bibitem{DMdeuterium1}
 C.~B.~Braeuninger and M.~Cirelli,
  Phys.\ Lett.\  B 678 (2009) 20
  [arXiv:0904.1165 [hep-ph]].
   
 \bibitem{DMdeuterium2}
 M.~Kadastik, M.~Raidal and A.~Strumia,
  arXiv:0908.1578 [hep-ph].
  
 \bibitem{pulsars}
A.~M.~Atoian, F.~A.~Aharonian and H.~J.~Volk,
Phys.\ Rev.\  D {52} (1995) 3265.
\hepart[0804.0220]{I. B\"ushing et al.}.
 T.~Kobayashi, Y.~Komori, K.~Yoshida and J.~Nishimura,  
Astrophys.\ J.\  {601} (2004) 340
[astro-ph/0308470]. See also the recent study in:
D.~Hooper, P.~Blasi and P.~D.~Serpico,
arXiv:0810.1527.
H.~Yuksel, M.~D.~Kistler and T.~Stanev,
  arXiv:0810.2784.
  S.~Profumo, arXiv:0812.4457. 
 See also P.~D.~Serpico,
arXiv:0810.4846  for an agnostic analysis.

 
 \bibitem{astro}
Tsvi Piran et al., arXiv:0902.0376. 
 
 
 \bibitem{PDG}
PDG: C. Amsler et al. (Particle Data Group), Physics Letters B667, 1 (2008)
 
 
 \bibitem{HisanoBBN}
  J.~Hisano et al.,
  arXiv:0901.3582.
  
  \bibitem{Jedamzik}
  K.~Jedamzik,
  Phys.\ Rev.\  D 70 (2004) 063524
  [arXiv:astro-ph/0402344].
  K.~Jedamzik,
  Phys.\ Rev.\  D 70 (2004) 083510
  [arXiv:astro-ph/0405583].
See also 
  F.~Iocco, G.~Mangano, G.~Miele, O.~Pisanti and P.~D.~Serpico,
  arXiv:0809.0631
  and references therein.
 
  
  \bibitem{Sommerfeld}
A. Sommerfeld, ``\"Uber die Beugung und Bremsung der Elektronen'', Ann. Phys. 403, 257 (1931).
J.~Hisano, S.~Matsumoto and M.~M.~Nojiri,
  Phys.\ Rev.\ Lett.\  {92} (2004) 031303
  [arXiv: hep-ph/0307216].
J.~Hisano, S.~Matsumoto, M.~M.~Nojiri and O.~Saito,
  Phys.\ Rev.\  D {71} (2005) 063528
  [arXiv: hep-ph/0412403].
See also previous work in 
  K.~Belotsky, D.~Fargion, M.~Khlopov and R.~V.~Konoplich,
  Phys.\ Atom.\ Nucl.\  {71} (2008) 147
  [arXiv:hep-ph/0411093] and references therein. 
The relevance of non-perturbative EW corrections
to DM freeze-out was pointed out in
J.~Hisano, S.~Matsumoto, M.~Nagai, O.~Saito and M.~Senami,
  Phys.\ Lett.\  B 646 (2007) 34
  [arXiv:hep-ph/0610249].

\bibitem{SomNonAb}
For a generic unbroken non-Abelian gauge interaction, the group theory needed to compute the Sommerfeld effect was presented in
A.~Strumia,
  Nucl.\ Phys.\  B  809 (2009) 308
  [arXiv:0806.1630 [hep-ph]].

\bibitem{Iengo}
R.~Iengo,
  arXiv:0902.0688 [hep-ph].
R.~Iengo,
  arXiv:0903.0317 [hep-ph].
   S.~Cassel, D.~M.~Ghilencea and G.~G.~Ross,
  arXiv:0903.1118 [hep-ph].
  
  \bibitem{singlet}
 V.~Silveira and A.~Zee,
  Phys.\ Lett.\  B {161} (1985) 136.
J.~McDonald,
  Phys.\ Rev.\  D {50} (1994) 3637
  [arXiv:hep-ph/0702143].
C.~P.~Burgess, M.~Pospelov and T.~ter Veldhuis,
  Nucl.\ Phys.\ B {619} (2001) 709 [hep-ph/0011335] and ref.s therein.

 \bibitem{FERMI} 
 \hepart[0905.0025]{FERMI/LAT collaboration}.
 
 \bibitem{HESS2009} 
 \hepart[0905.0105]{H.E.S.S. collaboration}.
  
\end{thebibliography}
\end{document}